\documentclass[12pt,preprint]{aastex}

\begin{document}

\title{The Disappearing Act of KH~15D: Photometric Results from 1995 to 2004} 

\author{Catrina M. Hamilton\altaffilmark{1}}
\affil{Astronomy Department, Mount Holyoke College, South Hadley, MA 
01075}
\email{chamilto@mtholyoke.edu}

\author{William Herbst}
\affil{Astronomy Department, Wesleyan University, Middletown, CT
06459}
\email{bill@astro.wesleyan.edu}

\author{Frederick J. Vrba}
\affil{US Naval Observatory, Flagstaff Station, Box 1149, Flagstaff, AZ
86002-1149} 
\email{fjv@nofs.navy.mil}

\author{Mansur A. Ibrahimov}
\affil{Ulugh Beg Astronomical Institute of the Uzbek Academy of Sciences,
Astronomicheskaya 33, 700052 Tashkent, Uzbekistan}
\email{mansur@astrin.uzsci.net}

\author{Reinhard Mundt and Coryn A. L. Bailer-Jones}
\affil{Max-Planck-Institut f\"{u}r Astronomie, K\"{o}nigstuhl 17,
D-69117 Heidelberg, Germany}
\email{mundt@mpia.de,calj@mpia.de}

\author{Alexei V. Filippenko and Weidong Li}
\affil{Department of Astronomy, University of California, Berkeley, CA 
94720-3411}
\email{alex@astro.berkeley.edu,wli@astro.berkeley.edu}

\author{V\'{\i}ctor J. S. B\'ejar}
\affil{GTC Project. Instituto de Astrof\'{\i}sica de Canarias, IAC. E-38200. 
La Laguna, Tenerife, Spain.}
\email{vbejar@ll.iac.es}

\author{P\'{e}ter \'{A}brah\'{a}m, M\'{a}ria Kun, Attila Mo\'{o}r,
J\'{o}zsef Benk\H o, and Szil\'{a}rd Csizmadia}
\affil{Konkoly Observatory, H-1525, P.O. Box 67, Budapest, Hungary}  
\email{csizmadi@konkoly.hu} 

\author{Darren L. DePoy, Richard W. Pogge, and Jennifer L. Marshall}
\affil{Department of Astronomy, Ohio State University, Columbus, OH 43210-1106}
\email{depoy@astronomy.ohio-state.edu, pogge@astronomy.ohio-state.edu,
marshall@astronomy.ohio-state.edu}

\altaffiltext{1}{Five College Astronomy Department, University of
Massachusetts, Amherst, MA 01003.}

\begin{abstract}

We present results from the most recent (2002--2004) observing campaigns of the 
eclipsing system KH~15D, in addition to re-reduced data obtained at Van Vleck 
Observatory (VVO) between 1995 and 2000.  
Phasing nine years of photometric data shows substantial
evolution in the width and depth of the eclipses.  The most recent data
indicate that  
the eclipses are now approximately 24 days in length, or half the orbital
period.  These results are interpreted and discussed in the context of the
recent models for this system put forward by Winn et al. (2004) and
Chiang \& Murray-Clay (2004).
A periodogram of the entire data set yields a highly 
significant peak at 48.37 $\pm$ 0.01 days, which is in accord with 
the spectroscopic 
period of 48.38 $\pm$ 0.01 days determined by Johnson et al. (2004).  
Another significant peak, at 9.6 days, was found in the periodogram of the
out-of-eclipse data at two different epochs.  We interpret this as the
rotation period of the visible star and argue that it may be tidally
locked in pseudosynchronism with its orbital motion.  If so, application
of Hut's (1981) theory implies that the eccentricity of the orbit is
$e$ = 0.65 $\pm$ 0.01.
Analysis of the UVES/VLT spectra obtained by 
Hamilton et al. (2003) shows that the $v\thinspace$sin$(i)$ of the visible 
star in this system is 6.9 $\pm$ 0.3 km s$^{-1}$.  Using this value of
$v\thinspace$sin$(i)$ and the measured rotation period of the star,
we calculate the lower limit on the radius to be 
$R = (1.3 \pm 0.1)\,$R$_{\odot}$, 
which concurs with the value obtained by Hamilton et al. (2001) 
from its luminosity and effective temperature.  Here we assume that 
$i$ = 90$\degr$ since it is likely that the spin and orbital angular 
momenta vectors are nearly aligned.
One unusually bright data
point obtained in the 1995/1996 observing season at VVO is interpreted 
as the point in time when the currently hidden star (B) made its last 
appearance.  Based on this datum, we show that star B is 0.46  
$\pm$ 0.03 mag brighter than the currently visible star A, which is entirely
consistent with the historical light curve (Johnson et al. 2005).
Finally, well-sampled $V_{J}$ and $I_{J}$ data obtained at the 
CTIO/Yale 1-m telescope during 2001/2002 show an entirely new feature: 
the system becomes bluer by a small
but significant amount in very steady fashion as it enters eclipse
and shows an analogous reddening as it emerges from eclipse.  
This suggests an extended zone of hot gas located close to, but above,
the photosphere of the currently visible star.  The persistance of
the bluing of the light curve shows that its length scale is comparable
to a stellar radius.

\end{abstract}


\section{Introduction}

KH~15D, an extraordinary pre-main-sequence eclipsing binary system, is 
located in the young (2--4 Myr) open cluster NGC 2264 ($d$ $\approx$ 760 pc).
In 1970, a Russian circular announced the discovery of eight new variable
stars in NGC 2264 (Badalian \& Erastova 1970).  KH~15D was among these, 
and was given the name SVS 1723.  It was found to vary by 1.1 mag and was
classified as an irregular variable.  In 1972, the new variables discovered
by Badalian \& Erastova (1970) were included in the 58th Name-List
of Variable Stars (Kukarkin et al. 1972), and KH~15D was given the
name V582 Mon.  Prior to October 2000, KH~15D had been observed in various 
surveys of the open cluster NGC 2264.  Flaccomio et al. (1999) and 
Park et al. (2000) both observed KH~15D at least once while the system was 
bright and included it in their catalogs, each identifying it as \#391 and 
\#150, respectively.
Earlier surveys of NGC 2264 (Herbig 1954; Walker 1956; Vasilevskis,  
Sanders, \& Balz 1965) did not identify KH~15D as a cluster member most 
likely because of its proximity to the B2~III star, HD~47887 (see Fig. 1).  
Its light was simply lost in the bright glow of the B2 III star.

This object gained its notoriety as KH~15D when it was reported to 
have unusual properties in 1995 based on a project at the Van Vleck 
Observatory (VVO) to photometrically monitor star-forming regions 
(Kearns \& Herbst 1998). 
This object was observed to undergo eclipses that were remarkably deep
($\sim$3.5 mag in $I$) and unusually long ($\sim$16 days in 1999/2000).
Of central interest was the evolution of the shape and duration of the 
eclipses that had been observed over a 5-year period (Hamilton et al. 2001). 
Out of eclipse, the observable star in this system was determined to have a 
spectral type of K6/K7 (Hamilton et al. 2001; Agol et al. 2004), while 
during eclipse, the observed spectrum was simply an attenuated version of 
that out of eclipse. 

An international campaign to monitor this intriguing system was begun 
during July 2001.  The primary goal of this 
project was to obtain much information about the structure 
of the intervening material.  Continued photometric observations from  
around the world provided nearly complete coverage of three consecutive 
eclipses during the 2001/2002 observing season (Herbst et al. 2002).  
These consecutive eclipses demonstrated that, in addition to the yearly 
changes observed in the phased data, differences in the shape of each  
individual eclipse could be observed over the course of a season.
The shapes of ingress and egress were successfully modeled by the steady
advance or retreat of a ``knife edge" across a limb-darkened star.
In order to match the duration of ingress and egress, however, Herbst
et al. (2002) determined that the occulting edge must be inclined by 
$\sim$15$\degr$ to the motion of the star. These authors
also reported that the object's color becomes slightly bluer 
during eclipse, which indicates that most (if not all) of the light
received during eclipse is due to scattering.

Polarization measurements made by Agol et al. (2004) and 
Gary Schmidt (private communication) show that out of eclipse, the system 
exhibits low polarization consistent with zero.  During
eclipse, however, the polarization is observed to increase 
to roughly 2\% across the optical spectrum.  These results support the 
conclusion that the star is likely completely eclipsed so that the flux 
during eclipse is due entirely to scattered radiation, and that the obscuring
material is most likely made up of relatively large grains ($\sim$10~$\mu$m; 
Agol et al. 2004).   
 
High-resolution spectra obtained during the December 2001 eclipse 
(Hamilton et al. 2003) revealed that the system is still undergoing 
accretion and driving a bipolar outflow.  Recently, extended molecular 
hydrogen emission near 2~$\mu$m was observed spectroscopically by Deming, 
Charbonneau, \& Harrington (2004).  
Given the spatial extent of the emission, and the observed H$_{2}$ line 
profile, Deming et al. (2004) suggest that the 
ambient H$_{2}$ gas is being shocked by a bipolar outflow from the star 
and/or disk.  Further evidence for an extended and well-collimated 
outflow was provided by Tokunaga et al. (2004), who obtained broad-band 
and narrow-band infrared images of KH~15D, showing a jet-like H$_{2}$ 
emission filament that extends out to $\sim 15\arcsec$ from the object.  
The position angle ($\sim$20\degr) of the in-eclipse polarization is 
nearly parallel to this H$_{2}$ filament (Agol et al. 2004).  

Many hypotheses have been proposed since 2001 to explain the evolving 
eclipses of KH~15D.  Most of these involve circumstellar material in 
some way.  Hamilton et al. (2001), Herbst et al. (2002), 
Winn et al. (2003), and Agol et al. (2004) all proposed the idea of an 
edge-on circumstellar disk with a warp that periodically passes in front 
of the star.  Barge \& Viton (2003) proposed that the eclipses were 
caused by an orbiting vortex of solid particles, and Grinin \& 
Tambotseva (2002) suggested that an asymmetric common envelope in a 
binary system could accurately reproduce the eclipses.  
Historical studies by Winn et al. (2003) found that observations made 
between 1913 and 1950 were consistent with no eclipses.  Johnson \& Winn 
(2004, hereafter JW04) discovered that between 1967 and 1982, the 
system alternated between bright and faint states with the same period 
as observed today, but 180$\degr$ out of phase with the current eclipses.  
Additionally, the eclipse depth was much smaller (only 0.67 $\pm$ 0.07 
mag in $I$ instead of 3.5 mag), and when out of eclipse, the system was 
brighter by 0.9 $\pm$ 0.15 mag than it is today (JW04).   
While Herbst et al. (2002) first suggested the possibility of a binary 
companion, the light curve between 1967 and 1982 led JW04 to believe that 
the light from a second brighter star once contributed to the flux coming 
from this system, and that the second star is now completely obscured.
These results have been confirmed and extended by data obtained at various
observatories between 1954 and 1997 (Johnson et al. 2005).  

Models of the KH~15D system that explained the historical and
modern-day light curves were put forward almost simultaneously by 
Winn et al. (2004) and Chiang \& Murray-Clay (2004).  Each model involves 
an eccentric pre-main-sequence binary whose motion periodically carries 
it behind a precessing circumbinary disk.  These models are supported by 
the results of a recent high-resolution spectroscopic monitoring program 
of KH~15D: Johnson et al. (2004) have observed radial-velocity variations 
that are consistent with a binary companion with an orbital period in 
agreement with the 48-day photometric period. 

In summary, KH~15D is now known to be a pre-main-sequence binary system
with a strongly eccentric orbit of 48.38 days.  The brighter
component B is currently totally obscured by the circumbinary ring or disk,
while only the fainter member A emerges from behind the obscuring
material for just less than half the period.  To explain the secular 
variations observed in the light curve over the past 9 years, it has 
been suggested that the eclipses occur whenever the motion of a star 
carries it behind the ring, with a period equal to that of the binary 
orbital period.  If this ring is also precessing, it would be possible 
to explain the changing length of the eclipse.

In this paper we present the analysis of photometric data
on KH~15D obtained at VVO between 1995 and 2000, color 
data obtained at the Cerro Tololo Inter-American Observatory (CTIO) during 
the 2001/2002 season, and the results of the 2002--2004 observing 
campaigns as compared to those of 2001/2002. 
Section 2 describes the principal sources of optical photometric
observations, the data reduction, and the resulting photometry for each 
participating observatory in the 2002--2004 campaigns.  
A comparison of individual data sets is shown in Section 3, while the 
results and analysis of these data are presented in Section 4.  A
brief discussion of our findings in light of the recent models is 
also presented, while a more complete interpretation is deferred to
a future paper (Winn et al. 2005, in preparation). 

\section{Observing Procedure, Reductions, and Photometry}

The principal sources of optical photometric data analyzed here 
are telescopes of 0.6--2.2~m aperture.  These 
include the US Naval Observatory's (USNO) Flagstaff Observing Station 
in Arizona, Mount Maidanak Observatory (MMO) in Uzbekistan, VVO
in Connecticut, the European Southern Observatory (ESO) and CTIO
in Chile, Kitt Peak National Observatory (KPNO) in Arizona, Wise 
Observatory in Israel, Teide Observatory in the Canary Islands, 
and Konkoly Observatory in Hungary. We also obtained data with
automated telescopes at the Tenagra Observatory in Nogales, Arizona, 
and the Katzman Automatic Imaging Telescope (KAIT), located at Lick 
Observatory atop Mt. Hamilton in California.  

While attempts were made to encourage uniformity in observing and 
reduction procedures within the initial group of participating 
observatories during the 2001/2002 season, the excitement that this 
object ignited inspired observers at telescopes around the world to 
contribute data to this project during the subsequent seasons.
Variations in seeing, image quality, sky brightness, flat-fielding
procedures, and CCD characteristics made it impossible to enforce 
substantial uniformity.  Although the photometry parameters varied widely 
between data sets for the 2002--2004 observing seasons, which are
discussed in detail here, we compared data obtained at different 
observatories on the same nights to search for consistency, as 
described below.  It should also be noted that not all observatories 
used the same reference stars in their data reduction.

We begin by listing the contributing observatories, telescopes, 
CCD parameters, filters, and dates in Table 1. (UT dates are used
throughout this paper.)  Most observations were made with a Cousins $I$ filter to 
reduce the nebular contamination of NGC 2264, as well as the contribution 
from the nearby B2~III star, HD~47887, which is $39\arcsec$ away.  
Here we will discuss the 2002--2004 data in detail.
Exposure times varied per observatory according to telescope and CCD 
pixel size.  Observatories with similar observing
procedures and reduction/photometry techniques are grouped together. 
Photometry parameters for each dataset from 2002--2004 are included 
in Table 2.

\subsection{USNO, Flagstaff Station}

\subsubsection{Observing Procedure and Reduction}

Attempts were made to observe KH~15D nearly every clear night during 
the 2002/2003 observing season at the USNO, Flagstaff Station.  
$B$, $V$, $R$, and $I$ filters were used while KH~15D was bright, and only
$V$, $R$, and $I$ during eclipse.
In 2003/2004,  multiple observations were obtained 
during three consecutive egresses using only the $V$, $R$,
and $I$ filters.  Exposure times generally ranged from 2 to 3 minutes, 
depending on whether the object was bright or faint.  All frames were 
bias-subtracted and flat-fielded in real time by local code.

\subsubsection{Photometry}
 
Photometry was performed on individual frames with an aperture radius that
depended on the full width at half-maximum (FWHM) of unblended stars in 
the image.  To maximize the signal-to-noise ratio (S/N), a radius of 
$\sim 1.5 \times {\rm FWHM}$ was employed.  
The background parameters were set such that the scattered light from 
HD~47887 was minimized (see Table 2).  

Differential photometry of KH~15D was performed using seven reference stars
selected from all-sky photometry obtained with the 1.0~m telescope at the 
USNO, Flagstaff Station.  A finding chart for the 7 comparison stars is 
shown in Figure 1, and the adopted standard Johnson $UBV$ and 
Kron-Cousins $RI$ magnitudes for each comparison star is given in Table 3.
To determine the magnitude for each frame ($B$, $V$, $R$, and $I$), the
$m_{true} - m_{instrumental}$ magnitude difference was calculated for each
non-saturated standard star.  This difference was then applied to the
$m_{instrumental}$ magnitude of KH~15D.  No color terms were used since the 
USNO filters are well matched to the standard system and the local standards
span the colors of KH~15D. The resultant uncertainty includes the 
spread in the $\Delta m$ due to not including a color term.

\subsection{Van Vleck Observatory (VVO), Tenagra Observatory, and KAIT} 

\subsubsection{Observing Procedure and Reduction}

KH~15D was generally observed once every clear night at VVO, Tenagra, and
KAIT.  Typically, five 1-minute exposures were obtained per
night; however, during ingress and egress, VVO increased the number of 
exposures.  All images obtained at VVO and 
Tenagra were bias-subtracted, corrected for dark current, and 
flat-fielded using standard IRAF\footnote
{Image Reduction and Analysis Facility, written and supported by 
the IRAF programming group at the National Optical Astronomy 
Observatories (NOAO) in Tucson, Arizona.  NOAO is operated
by the Association of Universities for Research in Astronomy (AURA), Inc.
under cooperative agreement with the National Science Foundation.} 
tasks (CCDPROC).  Images obtained at KAIT were bias-subtracted, corrected
for dark current, and flat-fielded automatically using a local code 
(Li et al. 2003).  After each image was processed, sets of five were 
averaged to increase the S/N, and the result was aligned to a 
single reference image using an IDL (Interactive Data Language) code.  
The FWHM for stars in each image was measured using the IRAF task 
IMEXAMINE, and was used to assess the seeing on that night.  Any image 
that exhibited a very large FWHM ($\sim$ 4.5$\arcsec$ or larger) was not used.

\subsubsection{Photometry}

For the analysis performed here, the average FWHM for the
observing season was calculated and used to 
determine the size of the
aperture radius utilized in the photometry process.  This method was chosen
so that all the data from one season could be processed and analyzed most
simply in batch mode.  We find that the standard deviation from this mean
is typically less than 0.5 pix.  To maximize the S/N,
a radius of $\sim 1.5~ \times <{\rm FWHM}>$ was chosen.

Instrumental magnitudes and uncertainties were computed for all the local 
comparison stars (see Fig. 1) and KH~15D using the IRAF task PHOT.
Calibration of the instrumental magnitudes from one 
observation to the next was accomplished by means of a set of comparison 
stars, which were averaged together to produce a synthetic stable 
comparison.
However, since even these stars are variable at some level,
the final set of stars used were identified by an iterative process, and 
represent the most stable ones, with a typical standard deviation of 
0.005 mag based 
on their night-to-night scatter.  Stars A, C, E, and G (see Fig. 1) 
were found to be most stable and were used as the comparison stars for 
both the 2002/2003 and 2003/2004 seasons at VVO.  Due to saturation issues, 
stars A, C, E, and F were used for the Tenagra images, while only 
stars C, E, and F could be used for the KAIT data due to the small size of 
the detector.
A time series of differential magnitudes relative to the synthetic comparison
star was then produced.  To place KH~15D on the standard magnitude scale,
the adopted $I$-band magnitudes (converted to flux) of each comparison star were
averaged to produce an $<I>$ comparison magnitude.  This was then added to 
the differential magnitudes calculated for KH~15D during the 2002/2003 and
2003/2004 observing seasons.

\subsection{Mount Maidanak Observatory (MMO)}

\subsubsection{Observing Procedure and Reduction}

KH~15D was observed multiple times on nearly every clear night at MMO 
(Uzbekistan) during the 2002--2004 observing seasons.  Exposure
times typically ranged from 2 to 5 minutes, and all observations were made
with a Bessell $I$ filter.  Raw frames were bias-subtracted and flat-fielded 
using the IMRED package within the IRAF environment.  Sky flat-field images 
were usually taken for each night, but in some cases only dome flats were 
available.  

\subsubsection{Photometry}

Differential magnitudes for KH~15D were computed in reference to star 16D
from Kearns et al. (1997).  In order to place these differential magnitudes
onto our standard scale, we needed to know the true magnitude of 16D. Thus,
16D was photometered during the 2001--2004 seasons at VVO in addition to the
7 local standards and KH~15D (see Section 2.2.2), and the weighted
average flux of 16D for the season was converted to a magnitude.  
This magnitude was then added to the differential magnitudes computed for 
15D at MMO.  The weighted average $I$ magnitudes computed for 16D 
between 2001 and 2004 are given in Table 4.  Since error bars were not 
produced when computing the differential magnitudes, we could not assign
formal photometric uncertainties to these data. 

\subsection{Konkoly Observatory}

\subsubsection{Observing Procedure and Reduction}

KH~15D was observed multiple times on nearly every clear night at 
Konkoly Observatory during the 2002/2003 observing seasons.  Exposure
times ranged from 2 to 20 minutes, but were typically 5 minutes.  All 
observations 
were made with a Cousins $I$ filter.  Raw frames were bias-subtracted and 
flat-fielded using the IMRED package within the IRAF environment.  
Sky flat-field images were usually taken for each night, but in some cases 
only dome flats were available.

\subsubsection{Photometry}

The differential magnitudes of KH~15D were measured in reference to
local secondary standard stars C, D, and F (see Fig. 1).
To determine the raw magnitudes of stars in the frames, aperture
photometry was performed using the IRAF/DAOPHOT package (Stetson 1990).
The FWHM of the stellar profile was determined for each night and an
aperture of $\sim 1.5 \times {\rm FWHM}$ was chosen.  
The ring from which we estimated the background contribution to 
the measured fluxes started 2 pixels from our aperture and had a width
of 5 pixels in every case.

The differential magnitudes were calculated by defining the variable's
brightness as $\Delta I = I_{var} - I_{comp}$.  These differential
magnitudes ($\Delta I$) were transformed onto the standard photometric 
system by taking the adopted $I$ magnitudes (converted to flux) of each
comparison star and averaging them together to produce an $<I>$
comparison magnitude.  This was then added to $\Delta I$ to get the 
true $I$ magnitude of KH~15D. No color term was applied 
since the Konkoly $I$-band filter is well matched
to the standard system.  Finally, three Cousins $I$ magnitudes of 
KH~15D were calculated on each frame by adding the magnitudes of the 
comparison stars to the differential magnitudes.  We calculated the
brightness by averaging the three different fluxes determined 
for KH~15D.  The uncertainty of the brightness of KH~15D was estimated
from the standard deviation of the three individual magnitudes.

\subsection{Teide Observatory}

\subsubsection{Observing Procedure and Reduction}

KH~15D was observed multiple times every clear night during the 2002/2003
observing season at the Teide Observatory.  Exposure times varied from 5 
to 20 minutes, and all observations were made with a Cousins $I$ filter.
Raw frames were bias-subtracted and flat-fielded using the IMRED package
within the IRAF environment.  Sky flat-field images were usually taken
for each night, but in some cases only dome flats were available.  

\subsubsection{Photometry}

Differential photometry of KH~15D was performed using an aperture of 
$0.5 \times {\rm FWHM}$.  The sky background was computed 
using a circular annulus
with an inner radius of $4 \times {\rm FWHM}$ and a width 
of 8 pixels.  
Differential photometry and calibration from instrumental magnitudes to
real ones was accomplished with respect to the reference stars 16D, 17D, 27D,
31D, and 39D from Kearns et al. (1997).

\subsection{Van Vleck Observatory Data from 1995 to 2000}

Data obtained at VVO between the years of 1995 and 2000 were re-examined
with parameters that were consistent with those of the 2001--2004 observing 
seasons.  This was done so that an absolute comparison between the 
earliest data and the most recent data could be made.  One difficulty 
in completing this task was the fact that from the fall of 1995 through 
the spring of 1998, VVO used a chip that was only 512 $\times$ 512 pixels in size.  
Therefore, only one of the current ``standard'' stars (D) was available, 
and it also happened to be the most variable star.  Therefore, we consulted the
original photometric studies of NGC 2264 conducted at VVO 
(Kearns et al. 1997; Kearns \& Herbst 1998) and examined 5 of their
quoted comparison stars (16D, 17D, 21D, 27D, and 31D).  Since these 
stars did not have known magnitudes, each was photometered on the images 
taken at 
VVO during the 2001--2004 observing seasons, allowing us to 
determine their 
magnitude on our standard scale.  Table 4 lists the weighted average $I$ 
magnitude and uncertainties for each star during 2001--2004.  Once a standard
magnitude was known for each comparison star, all 5 stars and
KH~15D were photometered on the images obtained at VVO between 
1995 and 1998.  Table 5 
lists the comparison stars used for each season. 

\subsection{European Southern Observatory (ESO)}

\subsubsection{Observing Procedure and Reduction}

The ESO observations of KH~15D contributed here were the byproduct of a larger 
photometric monitoring program of NGC 2264 (Lamm et al. 2004).  Full
details describing the observing procedure, reduction methods, and
photometry parameters can be found in Lamm et al. (2004).  In summary, 
monitoring of NGC 2264, and hence of KH~15D, took place during a period of 
two months between 30 December 2000 and 01 March 2001 with the Wide Field
Imager (WFI) on the MPG/ESO 2.2~m telescope on La Silla (Chile).  All 
observations were carried out with a Cousins $I$-band filter.  The WFI
consists of a mosaic of 4 $\times$ 2 CCDs with a total array size of 8K $\times$ 8K. 
Exposures of 5~s, 50~s, and 500~s were obtained, but the data used in this 
contribution come only from the 500~s images.  Image processing was done 
separately for
each of the individual WFI chips using standard IRAF tasks.  Each image
was bias-subtracted using the overscan region in each frame, and flattened
using an illumination-corrected dome flat.

The data presented here are based on differential photometry relative 
to a set
of non-variable reference stars.  The DAOPHOT/APPHOT task was used to measure
the brightness of KH~15D.  The aperture was chosen to be 8 pixels 
(1.9$\arcsec$)
in diameter for all measurements in order to maximize the S/N. 
The sky was calculated as the median of an annulus with an inner 
diameter of 30 pixels (7.1$\arcsec$) and a width of 8 pixels centered 
on KH~15D.
Because only differential magnitudes were calculated for the ESO data, 
we computed the average out-of-eclipse magnitude for the 2000/2001 season 
based only on the data obtained at VVO.  An appropriate out-of-eclipse
magnitude was then added to the ESO differential magnitudes such that these
data could be placed on our standard scale.

\subsection{Cerro Tololo Inter-American Observatory (CTIO)}

\subsubsection{Observing Procedure and Reduction}

KH~15D was observed extensively during the 2001/2002 observing season 
with the CTIO/Yale 1~m telescope. All observations
were made with a single instrument (ANDICAM; see DePoy et al. 2003 for a
description) using Johnson $V$ and $I$ filters.  
KH~15D was observed on nearly every usable night between 30 August 2001
and 18 April 2002.  Typically, several images were obtained each night 
using 300~s exposures.  The data were reduced in the usual manner using 
standard IRAF routines: an overscan (bias) was subtracted and a flat-field 
was applied to all the images.

\subsubsection{Photometry}

KH~15D was measured relative to two nearby stars shown in
Figure 2. Labeled as ``1" and ``2," they are also known as 5D and 
25D from Kearns et al. (1997), with $I$ = 13.250 $\pm$ 0.001 mag and $I$ = 16.464
$\pm$ 0.005 mag, respectively.  We find that neither comparison star
varied by more than 0.005 mag over the duration of the observations of
KH~15D.  A 3.6$\arcsec$ diameter aperture was synthesized on each of the CCD
images for both of the reference stars and KH~15D using standard IRAF
aperture photometry routines. The sky was determined in a 4$\arcsec$ to 
6$\arcsec$ annulus immediately adjacent to each of the sources.

We calibrated the measurements of KH~15D relative to the two stars using
images obtained at the 1.3-m McGraw-Hill telescope of the Michigan-Dartmouth-MIT (MDM)
Observatory.  The data were obtained on the photometric night of 19 October
2003.  The scale for the 1024 $\times$ 1024 pixel CCD that we used is 
0.5$\arcsec$ pixel$^{-1}$.  Throughout the night we observed photometric 
standard stars taken from Landolt (1992) at a range of airmasses.  
Near the end of this night we also
obtained 12 $V$-band and 7 $I$-band images of the KH~15D field.

We reduced the data using standard IRAF techniques, including 
bias-subtraction and flat-fielding using twilight-sky flats obtained during the
evening twilight of the same night.  We performed aperture photometry on the
McGraw-Hill data, using a 20 pixel (10$\arcsec$) diameter aperture for 
both the standard
stars and the target stars in the KH~15D field.  The seeing throughout the
night was roughly 1.4$\arcsec$, so aperture effects should be small. A
photometric solution was found using all 36 standard-star measurements in $V$
and $I$.  We solved only for the zero point and extinction coefficient terms
and held the color term constant at zero.  This
solution was then applied to the KH~15D field stars, and the calibration was 
used to convert the relative photometry to the standard system.
We estimate that ignoring the color term in our calibration observations
and between the MDM and the CTIO/Yale observations gives a systematic 
error in the KH~15D light curve of $\le$0.02 mag.  This is based on 
intercomparison of other standard-star measurements.

\section{Comparison of Photometric Data}

This paper combines a number of data sets obtained at various 
observatories, analyzed by different means, as described in the 
preceding sections.  Before beginning 
the process of analyzing the results of these data, random and 
systematic errors
between data sets were assessed.  To accomplish this, we selected all the 
photometry obtained at different places on the same Julian Date.  If there
were multiple observations made at a given observatory, a nightly mean was 
calculated.  This nightly mean was then used in the analysis described below.
Out-of-eclipse and in-eclipse observations were analyzed 
separately since the photometric errors are typically much larger during 
eclipse owing to the faintness of the object.  Additionally, times of 
ingress and egress were avoided since the magnitude of the system changes 
too rapidly for comparison of data, even as close as one day apart.  
Each season is addressed separately.

\subsection{The 2001/2002 Observing Season -- The CTIO Contribution}

One of the main objectives of the observing campaign of 2001/2002 was
to study, in detail, the eclipse of December 2001.  The original 
contributing observatories included Wise, VVO, MMO, USNO, and KPNO, and  
the results were reported by Herbst et al. (2002).  
We now have data from the CTIO/Yale 1~m telescope that has
provided us with perhaps the highest sampling rate yet of the eclipses 
that occurred during the 2001/2002 season.  These are shown in Figure 3.  
The central reversal is evident in these data, as well as some variation 
in the shape of ingress/egress from eclipse to eclipse.  Additionally,
substantial variability is seen outside of eclipse, as
explored further in Section 4.1.2.

Since these data were originally obtained through Johnson $V$ and $I$ 
filters, we could not immediately combine them with the rest of 
our Cousins $I$-band data.  The coverage of nearly 5 
eclipses, including dense sampling throughout ingress and egress, 
provided excellent color information about this system.  We therefore 
examined the Johnson colors separately from the other data sets and 
present the results in Section 4.  However, in order to add these 
Johnson data to the full $I$-band light curve,
we computed a transformation from Johnson $I$ to Cousins $I$ by comparing 
nearly simultaneous Johnson $V$-band and Cousins $I$-band data that were 
obtained at the USNO (see next section).

\subsection{The 2002/2003 Observing Season}

In 2002/2003, seven different observatories contributed to the light curve
of KH~15D (see Table 1).  To evaluate the systematic 
differences between each contribution, we began by selecting all the data 
that were obtained on the same Julian Dates outside of eclipse, for the
entire observing season.  We define
``out-of-eclipse" to correspond to phases between 0.3 and 0.7 for both the
2002/2003 and 2003/2004 seasons.  The out-of-eclipse 2002/2003 data 
are shown in Figure 4.  For each Julian Date, the mean brightness 
of the star 
was formed based on all measurements.  These contributions might include a 
single observation from one observatory and a mean observation (due to 
multiple exposures throughout the night) from another.  The difference 
between each observatory's measurement and the total nightly mean was then 
calculated and examined 
as a function of the total mean brightness.  A histogram indicating 
the number 
of times an observatory's measurement differed from the mean by 0.01 mag, 
0.02 mag, etc. is shown in Figure 5.  In general, there appears to be 
good agreement between the data sets.  Most observations differ from the 
mean by $\sim$0.01 mag, which is typically the size of the estimated 
random errors associated with the photometry.  Thus, we can say that the 
data 
sets are consistent with each other to within $\sim$0.02 mag.  Variations 
in the 
light curve on the order of $\sim$0.08 mag are observed out-of-eclipse 
(see Fig. 4), which we take to be real, and not an artifact of 
combining the data sets.  This will be explored further in Section 4. 

Figure 6 shows multiple data obtained on the same Julian Dates while in deep 
eclipse during 2002/2003.  We defined ``deep eclipse" to correspond to 
phases between 0.83 and 0.17 for both 2002/2003 and 2003/2004, with 
phase = 0.0 (and 1.0) representing the approximate phase for 
mid-eclipse.  The same analysis described above was performed on 
these data.  
Figure 7 shows that the differences of each observatory's observation from 
the mean is on average $\sim$0.15 mag, which is consistent with the 
photometric observational errors.

\subsection{The 2003/2004 Observing Season}

In 2003/2004, only four observatories contributed to the light curve 
of KH~15D (see Table 1).  There were significantly fewer multiple 
observations out-of-eclipse during 2003/2004 as compared to 2002/2003.  
We compared these individual measurements in the 
same manner as discussed above.  The agreement between
these data sets is not as good as was seen in the previous season.  
On average, the data from Tenagra were $\sim$0.04 mag brighter than 
those from VVO, and the data obtained at VVO were $\sim$0.02 mag brighter 
than those obtained at MMO (although that is not always the case).  
There are generally only 2 observations contributing to the calculation 
of the mean on any given night.  Therefore, the difference between each 
observatory's measurement and the mean gives us a sense of the systematic 
differences between each data set.  The average difference is 
$\sim$0.02 mag, which is slightly larger than the typical photometric 
errors ($\sim$0.01 mag or smaller).  The in-eclipse data from 2003/2004 
showed a great deal of scatter, but for the most part, appear to be 
correlated with  each other.  On average, the data depart from the 
calculated mean by $\sim$0.15 mag, which is on the order of the quoted 
photometric errors associated with the in-eclipse measurements.  

Overall, the data sets from both 2002/2003 and 2003/2004 mesh fairly 
well, and do not appear to be grossly different from each other.  
Therefore, we feel confident that the shape of the light curve and 
the depth of the eclipses are accurate to within $\sim$0.02 mag 
out-of-eclipse and $\sim$0.15 mag in deep eclipse.

\section{Results and Analysis}

The full data set now contains 6694 measurements of Cousins $I$ magnitude
covering nine seasons of observation between 1995 October and 2004 March, 
as well as additional color measurements. Magnitudes in $I$ are
given in Table 6 and colors are given in Table 7.
Both tables, in full, are available electronically.
In this section we describe the light and color variations and 
propose a rotation period for the star.  We also provide some qualitative
guidance to the interpretation of the variations based on the models of
Winn et al. (2004) and Chiang \& Murray-Clay (2004).  Detailed 
modeling of the photometric (and spectroscopic) data will be done by
Winn et al. (2005).  We turn first to a discussion of the light variations 
in $I$, where most of the data are available.

\subsection{Light Variations}

Figure 8 shows the data from the last two seasons,
which is the primary addition to the data set in this paper. Clearly,
we have obtained good coverage of a total of nine eclipses over 
two seasons.  Results indicate that the trends noticed by 
Hamilton et al. (2001) (reduction in amplitude of the central reversal
and widening of the eclipse) are continuing, and it would appear that 
the duration of the eclipse is now about one-half of the period 
(compare with Fig. 3).  The shapes of ingress and egress are not 
identical and show some level of variation, not only from eclipse 
to eclipse, but also from year to year.  Since 2001, the duration 
of the eclipse has been growing by roughly 2 days year$^{-1}$.  However, as 
will be seen in Section 4.1.2, this has not always been the case.  
In earlier years, such as from 1998/1999 to 1999/2000, the change in 
eclipse length was close to 1 day.

In Figure 9, the last two seasons are compared with the seven previous 
ones for which modern, CCD photometry is available. Some
of the features and a distinct secular variation of the light curve can
be seen.  Here, it appears that the out-of-eclipse brightness 
of the star has not changed by more than $\sim$0.1 mag since 1995, 
while the depth of the eclipse has grown by $\sim$1 mag over that same 
interval.  In fact, it appears as though the eclipse depth has been
increasing at a rate of $\sim$0.2 mag year$^{-1}$.  

To determine the best estimate of the mean $I$ magnitude 
out-of-eclipse, we looked solely at the data obtained at VVO since it 
was the only observatory from which we had data for every year.  We began by
calculating a mean $I$ magnitude per year to see if there was any 
variation in the out-of-eclipse brightness with time.  The phases that 
corresponded to the out-of-eclipse state changed as the width of the
eclipse evolved with time (discussed below), and are listed in Table 8. 
Figure 10 shows the mean $I$ magnitudes versus observing season at VVO.  
The straight line represents a simple linear fit to the data.  It
appears as if the currently visible star might also be fading a bit, but
we cannot be sure whether this is a significant trend.  
As shown below, the increasing depth of the
eclipse is accompanied by an increased time duration and a diminishing
height for the central brightness reversal. The current data show that
these trends, noticed previously (see Hamilton et al. 2001; 
Herbst et al. 2002), have continued in the last two seasons.

\subsubsection{Brightness of the Currently Invisible Star}

During the first season of monitoring at VVO there was one unusually
bright data point, taken on only the sixth night that the system was
observed. Repeated checks have shown that this is a valid datum and
cannot be discarded as an error. As will be shown below, its phase also
corresponds to a point close to mid-eclipse. We now interpret that
point as the only observation in our data set during which the
relatively unobscured light from the currently invisible star (B) was seen
directly.  This point occurred close to periastron when star B was 
peaking above the disk for just about the last time.  On that night, the
system's light was only due to star B, since star A was hidden behind the
disk.  If this interpretation is correct, it means that the
magnitude of star B is equal to or brighter than $I$ = 14.01 $\pm$ 0.01
(see Table 6).  Since
star B might also have been partly hidden itself, we cannot say for sure 
based on one measurement whether star B is brighter than $I$ = 14.01 
$\pm$ 0.01, but it certainly cannot be fainter.  

This conclusion is consistent
with the historical light curve.  Johnson et al. (2005) show that prior
to 1990, the system was at times as much as 1 mag brighter than it
appears today.  If we take the difference between the combined light 
(star A + star B) and the light of star A alone to be $-1.0$ mag, then the
magnitude difference between star A and B is 0.45.  The mean $I$ magnitude 
out-of-eclipse based on the VVO data (see Fig. 10) shows that star A has a 
mean $I$ = 14.47 $\pm$ 0.03 mag.  Assuming a minimum $I$ magnitude of 14.01 $\pm$
0.01 for star B, the mean brightness of star B is 0.46 $\pm$ 0.03 mag brighter 
than star A.   
To make further progress we wish to phase the light curve with the
appropriate period for the system, a task to which we now turn.

\subsubsection{The 48-day Period}

It is important to establish the period of this system to the highest
accuracy possible so that the data can be properly phased and the
evolution of the light curve with time correctly displayed. It is now
known that the 48-day period comes from the orbital motion within the
binary. As such, the most definitive determination of the period should
come from radial-velocity measurements and the spectroscopic-binary
solution. This yields a period of 48.38 days with an uncertainty of
about 0.01 days (Johnson et al. 2004). 
The problem with this method is that only a portion of the
full orbit of one star can currently be seen, and it is the part near
apastron which provides the least leverage on the solution. A second
complication is that the visible star may suffer from a Rossiter effect
(Worek 1996) of unknown size as it enters or emerges from eclipse, 
depending on the
inclination angle of the rotation axis to the line of sight. Therefore,
the data that provide the greatest leverage on the radial-velocity
solution (i.e., those taken during late egress or early ingress) may
also be the data which are compromised in this regard.

Hence, we believe it is important to ask independently of the radial
velocities whether the period of the system can be determined. The
problem with the photometric method is, of course, that the light curve
is evolving with time due to the progressive occultation by a
foreground screen. This puts an additional time-dependent behavior into
the system, albeit on a much longer time scale than the orbital period.

To proceed, we first did a Fourier transform of all of the modern
photometric data using the Scargle Periodogram technique (Scargle 1982; 
Horne \& Baliunas 1986). This yielded
a highly significant peak in the periodogram at 48.367 days (see Fig. 11), 
which we round to 48.37, as discussed below. This is gratifyingly close 
to the spectroscopic period and to the previously reported photometric 
period of 48.36 days (Herbst et al. 2002).

We attempted to refine the photometric period by focusing on a
particular feature in the light curve whose phase we believe is stable
and whose location we could determine from season to season. The only
such feature that exists is the peak of the central reversal (see Fig. 12), 
which we take to be the point in the orbit when the currently unseen star is
closest to the edge of the obscuring cloud. 
Assuming that this is a
stable, fixed location in the orbit, if we could phase these central
peaks we would have a good estimate of the period. Unfortunately, it
turns out that this method lacks discrimination because the location of
the central peak is not sufficiently well defined by the light curves,
especially the early ones where the data density is too low. A
difference of 0.01 days in period is not significant and does not affect 
our results, as it corresponds to only 0.75 days over the 75 cycles 
covered by our monitoring and the early data do not constrain
the time of the central peak to better than that.

Therefore, we find it impossible to improve on the accuracy of the
period determinations with just these data. It might be possible to do
this photometrically using the historical data, but any discussion of
that is postponed to a later contribution (Winn et al. 2005, in prep.). Here we
simply adopt the result of the periodogram analysis, 48.37 days,
noting that it is consistent with the spectroscopic determination; 
the 0.01 day difference does not affect any of the analysis or
conclusions reported here. 

\subsubsection{An Additional Periodicity -- The Rotation Period of the 
Visible Component}

During the 2001--2004 observing seasons, a great deal of effort was
made to observe KH~15D out-of-eclipse as well as in eclipse.
Significantly more data populated the bright phase of the light curve
during these seasons than ever before, and small-scale variations
($\sim$0.08 mag over the course of 8--14 days in 2002/2003) are 
evident (see Fig. 4 and Section 3.1).  Since the visible star in the 
KH~15D system is a weak-lined T~Tauri star (WTTS; Hamilton et al. 2001), 
it is most likely spotted in some fashion.  It has been determined that 
this K7 star is still actively accreting (Hamilton et al. 2003), and could 
therefore exhibit accretion hot spots on its surface, in addition to any dark, 
magnetically induced star spots.  If this is the case, a closer look at the 
out-of-eclipse data from the 2001--2004 seasons may provide an estimate for 
the rotation period of the star.

The analysis above indicates that systematic errors between data sets are 
relatively small, being at the level of 0.02 mag or less.  However,
when searching for the tiny amplitudes (less than 0.1 mag) that may 
characterize rotationally induced spotted-star variations, even errors of this
size may make the search more difficult.  Therefore, we have searched
each observatory's data set independently for periodicity during the 
out-of-eclipse phases.  It is also prudent to confine the search to a single
season of observation since spots are well known to evolve on time scales
of less than one year.  Hence, one rarely finds phase coherence of spot
variations extending over more than one observing season.

Detecting spots on WTTSs that are not undergoing eclipses is difficult
enough, and generally only 15\% or less of the stars in a cluster field will
exhibit coherent periodic variations over a season's observing. In this
star we have greater difficulty because during about half the observing
season the star is in eclipse. Nonetheless, we have found that
periodogram analyses of the available data sets reveals highly
significant periods in three cases (see Fig. 13), as listed in 
Table 9. Notably, two of these periods are identical to within the 
errors, averaging to 9.6 days. The chance of finding two exactly equal, 
highly significant periods in these data is vanishingly
small if the periodicity is not real; we thus identify the
detected period as the likely rotation period of the visible star. It
is interesting that the third significant period found in the
periodograms (7.95 d) is very close to the beat period between 9.6 and
48.37 days.

Supporting evidence for our interpretation comes from the measured 
$v\thinspace$sin$(i)$ of the star. As discussed in Appendix A, we derive 
a value of $v\thinspace$sin$(i)$ = 6.9 $\pm$ 0.3 km s$^{-1}$ for the visible 
component based on high-resolution spectra obtained out of eclipse. Since 
sin$(i)$ is likely to be
very close to 1.0 for this star (assuming that the spin and orbital
angular momenta vectors are nearly aligned), one can easily calculate
that for a radius of $R = (1.3 \pm 0.1)\,$R$_{\odot}$ (Hamilton et al. 2001), 
the expected
rotation period is 9.6 $\pm$ 0.1 days, consistent with our result. Both
the photometry and spectroscopy of this system out-of-eclipse concur in
suggesting a rotation period of 9.6 days for the visible component. 
To within the errors, this is a 5:1 resonance with the orbital period
(48.37/5 = 9.67), although it is not clear whether this is physically 
significant or just a numerical coincidence.

A rotation period of 9.6 days is somewhat long for a 0.6~M$_{\odot}$ WTTS
in NGC 2264, although not unprecedented.  The (bi)modal values for stars
more massive than 0.25~M$_{\odot}$ are around 1 and 4 days.  However,
about 12\% of the 182 cluster stars in that mass range with known 
rotational periods have $P > 9.6$ days (Lamm et al. 2005).  Could the 
slightly slower than typical rotation of the visible star in the KH~15D 
system be caused by a tidal interaction between the components?  The 
effect of tidal friction on rotation can be estimated quantitatively
based on the seminal theory of Zahn (1977) for stars with convective
envelopes.  He shows that the synchronization time for a star's rotation
to become tidally locked to its revolution is approximately given by
$$t_{synch} = {q^{-2} ({a \over R})^6},$$ where $q$ is the mass ratio, $a$ 
is the semimajor axis, $R$ is the stellar radius, and $t_{synch}$ is given 
in years. Most of the torque in a highly eccentric system will naturally 
occur at periastron. With $q \approx 1$ and $a/R = 13$, we find a 
synchronization 
time of $\sim$5 My. Hence, it is not unreasonable to suggest that 
tidal friction is responsible for some degree of slowing of this WTTS.

We further note that since the orbit is rather eccentric, one would not  
expect to reach synchronous rotation, but only pseudosynchronous  
rotation.  The pseudosynchronization timescale describes a near 
synchronization of revolution and rotation around periastron (Hut 1981). 
Hut (1981) has shown in elegant fashion that for binaries  
with eccentricities exceeding $\sim$0.3, as must surely be the case for  
KH~15D, tidal interaction near periastron is expected to produce an  
equilibrium angular rotation velocity of about 0.8 times the orbital  
angular velocity at periastron. One may write, therefore, that the  
pseudosynchronous rotation period ($P_{ps}$) predicted by this theory is  
related to the orbital period ($P_{orb}$) by
$$P_{ps} = {{P_{orb} \over f}{{{(1-e^2)}^{3 \over 2}} \over  
{{(1+e)}^2}}},$$
where $f$ is near 0.8 for $e > 0.3$ and is given precisely by Hut (1981).

The range of particular interest for KH~15D is $0.68 < e < 0.8$ as  
discussed by Johnson et al. (2004), because these solutions are consistent 
with the radial velocities and with the inferred masses of the components  
based on astrophysical constraints. Over this limited range, and  
somewhat beyond it, $f$ = 0.81 $\pm$ 0.01. Adopting $P_{ps} = 9.6 \pm 0.1$  
and $P_{orb}$ = 48.37 $\pm$ 0.01 yields a rather precise prediction  
for $e$ of 0.65 $\pm$ 0.01. This is the required orbital eccentricity if  
the star is in pseudosynchronous rotation and if the theory of  
Hut (1981) is correct. Remarkably, this eccentricity is very close to  
those inferred on the basis of two independent methods. As already  
noted, the orbital solution based on radial velocities, together with  
astrophysical constraints on the masses of the components, leads to $0.68  
< e < 0.8$, just barely outside of the present result. Model fits to  
the light curve and its evolution over more than fifty years, by  
Winn et al. (2004) and Winn et al. (2005, in prep.), also suggest an 
eccentricity of between 0.55 and 0.7.

Since the rotation rate of the visible component is notably slower than  
that of most stars of comparable mass and age, and since it agrees so well 
with the predicted pseudosynchronous period, we suggest that the star has,  
indeed, been tidally locked into its current configuration. That, in  
turn, puts a rather severe limit on the possible eccentricity of the  
orbit, assuming that the locking has reached equilibrium and that the  
theory of Hut (1981) is accurate in its prediction of $f$. We note that  
the strong sensitivity of $P_{ps}$ to $e$ means that there is only a 
very small range of measured rotation periods which would have given  
consistent results with the orbital solution and light-curve fitting  
techniques. For comparison, a normal rotation period of around 1--4 days  
for this star would have predicted a value of $e$ = 0.8--0.95 for this  
binary, rather extreme solutions that are inconsistent with the other  
results.

\subsection{The Color Behavior}

While most of the monitoring has been in the $I$ band due to the brightness
of the star during eclipse, we have also obtained some data at other 
wavelengths
which reveal interesting features of the system.  Here we discuss the color
results obtained at CTIO in 2001/2002 and at the USNO during several 
seasons.  The discovery reported by Herbst et al. (2002) that the star is
generally bluer when fainter is confirmed, but we now have much more 
detailed and intriguing information on the color variations and their
phase dependence.

\subsubsection{CTIO Observations}

Extensive $V$-band and $I_{J}$-band observations were made during the 
2001/2002 season at CTIO with the Yale 1~m telescope (see Sections 2.8
and 3.1 for details) and are shown in Figures 14 and 15.  
The solid and dashed lines in Figure 14 are provided to indicate the 
approximate beginning of ingress and ending of egress, respectively.  While
the Johnson magnitudes can be transformed to Cousins for comparison
with other data sets, there is no need to do so in this section, so we plot
the results in the Johnson system. 

The high density of color data obtained at CTIO during this season reveals
an interesting new feature of the behavior of KH~15D: it becomes bluer by a
small but significant amount in very steady fashion as it enters eclipse
and shows an analogous reddening as it emerges from eclipse.  This is
quite easily seen in both the time and phased versions of the light curves
(Fig. 14 and 15, respectively).  It is also confirmed by the less dense
but more temporally extended color monitoring done at the USNO as described
in the next section.

While the general trend of bluer when fainter had been recognized for
this star previously (Herbst et al. 2002) and has been seen in other
young stars, such as UXors (e.g., Herbst \& Shevchenko 1998), the smooth
decrease (increase) in $V-I$ seen during ingress (egress) in Figure 15 is
remarkable and unprecedented.  Before discussing its interpretation, we
complete a description of the color behavior during eclipse.

As both Figures 14 and 15 show, the color change that occurs as the star
goes into eclipse does not continue smoothly throughout the eclipse.  While
the data are noisy due to the low brightness of the star at those phases,
it is clear that the star does not maintain a steady color with phase during
eclipse.  Our impression of Figure 15 is that there are actually three 
phases when the star has its bluest colors, peaking at about 
$V-I_{J}$ = 1.2--1.3 mag,
or 0.3--0.4 mag bluer than the out-of-eclipse value of $V-I_{J}$ = 1.6 mag.  
These phases occur around $\pm$0.17 and close to phase 0.

It is interesting that there is also a distinctive feature in the light curve
at each of these three phases.  At $\pm$0.17 phase (in 2001/2002) there is a 
distinct change of slope in the decline rate during ingress and the rise rate
during egress.  While this distinctive change of slope can be seen in the
2001/2002 light curve, it is even more noticeable in the light curves of 
the last two years (see Fig. 8), where it occurs at phases of about $\pm$0.20 
and $\pm$0.24, respectively.  The other feature in the light curve
which seems to correspond to a blue extreme in the colors is the well-known
central reversal near phase 0.  Before discussing the interpretation of these
interesting features, we turn to a description of the other major set of
color data, that obtained at the USNO.

\subsubsection{USNO, Flagstaff Station Observations}
  
$V$, $R_{C}$, and $I_{C}$-band monitoring took place at the USNO 
1.3~m telescope throughout the 2001/2002 season, while $B$, $V$, $R_{C}$, and 
$I_{C}$-band monitoring took place at the 1.0~m telescope 
during 2002/2003 (see Section 2.1 for details).  
It was determined from the first observing season that the object was 
simply too faint in $B$ to provide any reliable photometry at those 
wavelengths during eclipse.  Over the second season of monitoring, 
$B$-band measurements were obtained out-of-eclipse, in addition to the 
standard $V$, $R_{C}$, and $I_{C}$ observations.  During 2003/2004, only 
$V$, $R_{C}$, and $I_{C}$ data were obtained during three consecutive 
egresses.

In Figure 16, we show the behavior of all of the Cousins-band colors
monitored at the USNO during the 2002/2003 season.  The trend is clearly
quite similar to what was seen in the CTIO data in 2001/2002 --- namely,
there is a distinctive bluing of the system just as it enters eclipse
or corresponding reddening as it emerges from eclipse.  In addition, there
are clearly some significant color variations out-of-eclipse which we
attribute to star-spot activity on the visible star.

Combining the color data obtained over many seasons at the USNO, we
illustrate the overall trend of color with brightness in Figure 17 .  This
confirms our impression that, out of eclipse, the star is about as red as
it gets and shows relatively little color variation beyond what may be
explained through star spots.  As it enters (emerges from) eclipse the star
gets bluer (redder) by 0.1--0.2 mag.  Near minimum light there are
substantial color variations; the star is sometimes at its bluest 
extreme, which is 0.3--0.4 mag bluer in $V-I$ than out-of-eclipse.  However,
sometimes it shows very little, if any, color effect.
 
\subsubsection{Interpretation of the Color Data}

A full explanation of the color data must await improved understanding
of the details of this system, which will only come from an extensive
phenomenological model currently being constructed (Winn et al. 2005, in 
prep). However, qualitative explanations for the observed trends can 
be proposed based on a few simple arguments. These lead to a set of
questions that need to be addressed, and they suggest further observations
that may clarify the situation.

Basically, blue colors indicate either the importance of
wavelength-dependent scattering, such as from small grains or
molecules, or the presence of a hotter component in the system, or
both. It is likely that there is a hotter component in the system,
since the currently invisible companion to the K7 star is known to be
more luminous, as discussed above. All models of pre-main-sequence
contraction indicate that the more massive star in a coeval pair is
also more luminous and hotter, so it is reasonable to suppose that the
currently invisible component of this system is also bluer than the K7
star. However, since the system was never substantially brighter in the
past, even when both stars were visible, it is also true that this
invisible component cannot be much more massive, luminous, or hotter than
the K7 star. In fact, the lack of detectable change of spectral type
during minimum means there is little other than K6 or K7 light in the
system, as far as a stellar component.

The original interpretation proposed for the color variation by Herbst
et al. (2002) is that near minimum light, we see the system primarily
or only by reflected light and that some small grains are involved in
the reflection. This is supported by the increased polarization
detected near minimum light by Agol et al. (2004). The current data
show, however, that there is almost certainly not a single color to
which the star moves during eclipse, but rather that there is a great 
deal of real variability, some or all of which is phase dependent. Therefore, 
it cannot simply be that all of the light of the system is coming from
distant, scattered radiation.

In particular, we are impressed by the relative peak in blueness that
occurs near phase zero in the color-phase plots (Fig. 15 and 18). This
is the time when the hotter, currently invisible, star is closest to
the edge of the occulting cloud. We propose that the blue peak at this
phase may be caused by this fact. We are apparently seeing a very small
amount of either transmitted light or increased importance in the
reflected component due to the light of this star, or, perhaps, the
extension of a hot, circumstellar nebula associated with it, above the
obscuring wall. In any event, we tentatively attribute the central peak
in blue color and the phase-dependent variations around it as due to
the orbital motion of a slightly bluer component (than the K7 star)
relative to the edge of the occulting disk and, perhaps, relative to
the principal scatterers.

The blue peaks associated with phases $\pm$0.17 in 2001/2002 are an
entirely new feature of the light of this system revealed by the
intensive color monitoring at CTIO during that season and deserve some
explanation. We find it interesting, as noted above, that they seem to
correlate with a distinct change in slope that occurs in the decline
(rise) in magnitude during late ingress (early egress).

One problem in interpreting these data is that we do not yet know how
sharp the occulting edge of the cloud is. Let us suppose, for the
purposes of this qualitative study, that the edge is very sharp. In that
case, we can identify the change of slope in decline (rise) rate as the
point where the photosphere has just been completely covered (has first
appeared). The fact that the color peaks blue at that point indicates
that the blue light is closely associated with the photosphere, arising
just above it. We suggest, therefore, that the light in this system
contains a hotter component associated with an extended chromosphere or
corona of the K7 star (and possibly its invisible companion as well).
We have no idea whether the physical nature of this is a scaled-up 
solar-like chromosphere, a magnetically channeled accretion column, 
or something else. But the color evidence suggests an extended
hotter, dense, optically thick zone located close to the photosphere of
the K7 star. The persistence of the bluing of the light curve shows
that it must be extended on a length scale comparable to a stellar
radius.

An alternative model is that one is seeing a peculiar scattering
effect just as the star enters (emerges) from full eclipse. One could
imagine, for example, that strongly forward scattered light might
glance off the top of an occulting disk providing a bluish ``glint" at
these phases. This would require, of course, wavelength-dependent
scattering and, therefore, some component other than the obscuring
cloud (which produces no reddening of the transmitted light). If
wavelength-dependent reflection is heavily involved here, then there
should be some dramatic increases in polarization during these phases.
It would definitely be interesting to extend the polarization study of
Agol et al. (2004) to a wider variety of phases.  It will also be
important to continue to monitor the color during these phases in
future eclipses, hopefully including the $U$ band with larger telescopes, 
which will give important information on the nature of the scattering 
or the temperature of the emission zone.

\section{Conclusions}

The light curve of the remarkable system KH~15D continues to evolve, evidently 
as a result of an obscuring screen moving across the orbit of a binary.  A 
periodogram analysis of our data indicates that the photometric period is 
in excellent agreement with the spectroscopic period.  Analysis of the
data collected outside of eclipse suggests that the visible star has a
rotation period of $\sim$9.6 days. This conclusion is in accord with the predicted
rotation period, which is derived from the measured $v\thinspace$sin$(i)$
(6.9 km s$^{-1}$) and taking sin$(i)$ $\approx$ 1.  

Interpretation of the color information leads us to 
believe that some extended blue emission is revealed in the colors as 
the currently visible star goes into or comes out of eclipse.  The eclipse 
length is now greater than half the orbital period of the system and 
continues to evolve, expanding at a rate that is $\sim$2 days year$^{-1}$.  Continued 
photometric monitoring will help to further constrain the models.  This 
object will clearly provide us with the opportunity to learn about disk 
evolution, the circumstellar environment of young stars, and the 
possibility of planet formation.  Thus, we strongly encourage 
continued monitoring before it disappears!

\acknowledgements

We would like to thank Josh Winn for helpful discussions regarding
this system.
C. M. H. thanks Christopher Johns-Krull for supporting her during
the last year of this project and providing the K7 template used in the
determination of $v\thinspace$sin$(i)$.
W. H. gratefully acknowledges support by the National Aeronautics and 
Space Administration under Grant NAG5-12502, issued through the Origins 
of Solar Systems Program.
P. A., M. K., A. M., J. B., and S. C. are supported by the Hungarian
National Science Fund under numbers T043739 and T043504.
A. V. F. and W. L. are supported by NSF grant AST-0307894.
A. V. F. is also grateful for a Miller Research Professorship at UC 
Berkeley, during which part of this work was completed.
V. J. S. B. would like to acknowledge Prof. Rafael Rebolo for informing
him of the international campaign to monitor KH~15D and for inviting him 
to participate.
KAIT was made possible by generous donations from Sun Microsystems, Inc.,
the Hewlett-Packard Company, AutoScope Corporation, Lick Observatory,
the National Science Foundation, the University of California, and the
Sylvia \& Jim Katzman Foundation.

\appendix
\section{Appendix: $v\thinspace$sin$(i)$ Results} 

In order to determine the $v\thinspace$sin$(i)$ of the visible star in the
KH~15D system, the UVES/VLT spectra (Hamilton et al. 2003) were revisited.
First, a K7~V model spectrum was produced using a temperature of 4000~K, a
log($g$) = 3.5 ($g$ in cgs units), and a macroturbulence value of 0.  The template 
was then artificially broadened to resemble that of a higher velocity star by
convolving it with a rotation profile that was generated from a theoretical
model, and including a macroturbulence value of 2 km s$^{-1}$.  This was accomplished 
by using an IDL code and various values of $v\thinspace$sin$(i)$ believed 
to span the range within which the $v\thinspace$sin$(i)$ of the visible 
K7 star is expected to be found.

Each broadened spectrum was then cross-correlated against the original
narrow-lined template, and the FWHM of the peak of the cross-correlation
function was measured.  A plot of the FWHM values versus 
$v\thinspace$sin$(i)$ -- a calibration curve -- was produced by fitting
a second-order polynomial function to the data.
The 29 Nov. 2001 spectrum was then
cross-correlated against the narrow-lined template and the FWHM value of
the cross-correlation peak was measured.  This value was determined to be
0.489.  The polynomial fit to the calibration curve was evaluated
at this FWHM value to obtain a $v\thinspace$sin$(i)$ = 6.9 $\pm$ 0.3 km s$^{-1}$ 
for the K7 star.  The uncertainty was determined by first broadening our
template spectrum to the value of 6.9 km s$^{-1}$.  We then performed a Monte
Carlo test where noise, appropriate to that which is found in the UVES/VLT 
spectrum, was added to the broadened template.  This broadened, 
noisy template was then cross-correlated against the narrow-lined 
model spectrum.  The FWHM of the peak
of the cross-correlation function was measured for each test, and the 
standard deviation of these values provided us with our estimated 
uncertainty.  


\clearpage
\begin{deluxetable}{lclcl} 
\tablewidth{0pt}
\tablecaption{Contributing Observatories for the 1995--2004 Seasons. 
\label{tbl-1}} 
\tablehead{
\colhead{Observatory} & \colhead{Aperture} & \colhead{CCD Parameters} &
\colhead{Filters} & \colhead{Dates Observed (UT)} } 

\startdata

VVO & 0.6 m & 0.5K $\times$ 0.5K, 0.6$\arcsec$ pix$^{-1}$ & $I$ & 27 Oct 1995 -- 25 Mar 1998\\ 

VVO & 0.6 m & 1K $\times$ 1K, 0.6$\arcsec$ pix$^{-1}$ & $I$ & 10 Dec 1998 -- 16 Mar 2004\\

ESO & 2.2 m & 8K $\times$ 8K, 0.238$\arcsec$ pix$^{-1}$ & $I$ & 30 Dec 2000 -- 1 Mar 2001\\ 

CTIO & 1.0 m & 2K $\times$ 2K, 0.3$\arcsec$ pix$^{-1}$ & $V,I_{J}$ & 22 Aug 2001 -- 18 Apr 2002\\ 

MMO & 1.5 m & 2K $\times$ 0.8K, 0.26$\arcsec$ pix$^{-1}$ & $I$ & 12 Sept 2001 -- 18 Mar 2004\\ 

USNO & 1.3 m & 2K $\times$ 4K, 0.6$\arcsec$ pix$^{-1}$ & $V,R,I$ & 28 Nov 2001 -- 3 Apr 2002\\

USNO & 1.0 m & 1K $\times$ 1K\tablenotemark{a}, 0.68$\arcsec$ pix$^{-1}$ & $B,V,R,I$\tablenotemark{b} & 11 Nov 2002 -- 30 Apr 2003\\ 

USNO & 1.0 m & 1K $\times$ 1K, 0.68$\arcsec$ pix$^{-1}$ & $V,R,I$ & 23 Dec 2003 -- 12 Mar 2004\tablenotemark{c}\\

KPNO & 0.9 m & 2K $\times$ 2K, 0.6$\arcsec$ pix$^{-1}$ & $U,B,V,R,I$ & 1 -- 22 Dec 2001\\

WISE & 1.0 m & 1K $\times$ 1K, 0.7$\arcsec$ pix$^{-1}$ & $I$ & 2 -- 22 Dec 2001\\ 

Tenagra & 0.81 m & 1K $\times$ 1K, 0.87$\arcsec$ pix$^{-1}$ & $I$ & 2 Nov 2002 -- 4 Apr 2004\\ 

KAIT & 0.76 m & 0.5K $\times$ 0.5K, 0.8$\arcsec$ pix$^{-1}$ & $I$ & 20 Sept 2002 -- 30 Mar 2003\\ 

Teide & 0.82 m & 1K $\times$ 1K, 0.432$\arcsec$ pix$^{-1}$ & $I$ & 28 Sept 2002 -- 28 Mar 2003\\ 

Konkoly & 1.0 m & 1K $\times$ 1K, 0.33$\arcsec$ pix$^{-1}$ & $I$ & 1 Oct 2002 -- 2 Mar 2003\\ 

\enddata

\tablenotetext{a}{A 2K $\times$ 2K CCD was also used.  It has the same pixel scale
as the 1K $\times$ 1K CCD chip.}

\tablenotetext{b}{Only the $V$, $R$, and $I$ filters were used during eclipse.}

\tablenotetext{c}{$V$, $R$, and $I$ observations were only made during egress.}

\end{deluxetable}

\clearpage
\begin{deluxetable}{llll}  
\tablewidth{0pt}
\tablecaption{Photometry Parameters for the 2002--2004 Seasons. \label{tbl-2}}
\tablehead{
\colhead{Observatory} & {Aperture Radius} & {Inner Sky Radius} & {Width}\\
\colhead{} & \colhead{in Pixels} & \colhead{in Pixels} & \colhead{in Pixels}}  

\startdata

VVO & 7 (4.2$\arcsec$) & 10 (6.0$\arcsec$) & 5 (3.0$\arcsec$)\\

MMO & 10 (2.6$\arcsec$) & 11 (2.86$\arcsec$) & 5 (1.3$\arcsec$)\\

USNO & 4.5\tablenotemark{a} (3.06$\arcsec$) & 6.5 (4.42$\arcsec$) & 2 (1.36$\arcsec$)\\

Tenagra & 3 (2.61$\arcsec$) & 5 (4.35$\arcsec$) & 2 (1.74$\arcsec$)\\

KAIT & 3 (2.4$\arcsec$) & 5 (3.2$\arcsec$) & 2 (1.6$\arcsec$)\\

Teide & 2.5 (1.1$\arcsec$) & 20 (8.6$\arcsec$) & 8 (3.5$\arcsec$) \\

Konkoly & 12 (3.96$\arcsec$)\tablenotemark{b} & 15 (4.95$\arcsec$)\tablenotemark{b} & 5 (1.65$\arcsec$) \\

\enddata

\tablenotetext{a}{The number 4.5 listed here is a weighted average
of the actual apertures used.  A 4-pixel radius was used 60\% of the time,
a 5-pixel radius was used 30\% of the time, and a 6--7-pixel radius was used
10\% of the time.  The largest radius was used during bad seeing, and only
when the object was bright.}

\tablenotetext{b}{This is the average value.  For the worst seeing, we used
18 and 23 pixels for the aperture and inner sky radius, respectively.}

\end{deluxetable}

\clearpage
\begin{deluxetable}{cccccc} 
\tablecaption{Adopted $UBVRI$ magnitudes and 1$\sigma$ uncertainties for the 
7 Comparison Stars. \label{tbl-3}}
\tablewidth{0pt}
\tablehead{

\colhead{Star} & \colhead{$U$} & \colhead{$B$} & \colhead{$V$} & \colhead{$R$} &\colhead{$I$} }

\startdata 

A & 14.401 $\pm$ 0.022 & 14.115 $\pm$ 0.021 & 13.368 $\pm$ 0.020 & 12.952 $\pm$ 0.020 & 12.548 $\pm$ 0.021\\

B & 13.514 $\pm$ 0.022 & 13.341 $\pm$ 0.021 & 12.624 $\pm$ 0.020 & 12.196 $\pm$ 0.020 & 11.734 $\pm$ 0.021\\

C & 13.563 $\pm$ 0.022 & 13.566 $\pm$ 0.021 & 12.969 $\pm$ 0.020 & 12.617 $\pm$ 0.020 & 12.240 $\pm$ 0.021\\

D & 15.808 $\pm$ 0.028 & 15.921 $\pm$ 0.022 & 15.013 $\pm$ 0.020 & 14.328 $\pm$ 0.021 & 13.645 $\pm$ 0.021\\

E & 15.803 $\pm$ 0.028 & 15.004 $\pm$ 0.022 & 13.885 $\pm$ 0.020 & 13.211 $\pm$ 0.021 & 12.614 $\pm$ 0.021\\

F & 15.319 $\pm$ 0.024 & 14.745 $\pm$ 0.021 & 13.869 $\pm$ 0.020 & 13.375 $\pm$ 0.021 & 12.902 $\pm$ 0.021\\

G & 16.842 $\pm$ 0.041 & 15.642 $\pm$ 0.022 & 14.230 $\pm$ 0.020 & 13.350 $\pm$ 0.021 & 12.479 $\pm$ 0.021\\ \hline

\enddata 

\end{deluxetable}

\clearpage
\begin{deluxetable}{llll} 
\tablecaption{$I$ Magnitudes and 1$\sigma$ uncertainties for Additional 
Comparison Stars. \label{tbl-4}}
\tablewidth{0pt}
\tablehead{
\colhead{} & \colhead{2001/2002} & \colhead{2002/2003} & \colhead{2003/2004} } 

\startdata

\bf{16D} & 14.303 $\pm$ 0.001 & 14.304 $\pm$ 0.001 & 14.302 $\pm$ 0.002 \\
\bf{17D} & 15.085 $\pm$ 0.001 & 15.098 $\pm$ 0.002 & 15.158 $\pm$ 0.003 \\
\bf{21D} & 15.023 $\pm$ 0.001 & 14.991 $\pm$ 0.002 & 15.006 $\pm$ 0.003 \\
\bf{27D} & 13.783 $\pm$ 0.001 & 13.785 $\pm$ 0.001 & 13.794 $\pm$ 0.002 \\
\bf{31D} & 14.423 $\pm$ 0.001 & 14.426 $\pm$ 0.001 & 14.411 $\pm$ 0.002 \\
 
\enddata

\end{deluxetable}

\clearpage
\begin{deluxetable}{cc}
\tablecaption{Comparison Stars for 1995--2000 VVO Data. \label{tbl-5}}
\tablewidth{0pt}
\tablehead{
\colhead{Observing Season} & \colhead{Comparison Stars} } 

\startdata

1995/1996 & 16D, 31D \\
1996/1997 & 16D, 31D \\
1997/1998 & 27D, 31D \\
1998/1999 & A, B, C, F \\
1999/2000 & A, B, C, F, G \\
2000/2001 & A, B, C, F \\

\enddata

\end{deluxetable}

\clearpage
\begin{deluxetable}{cccc}
\tablecaption{Photometric Measurements of KH~15D.\tablenotemark{a} 
\label{tbl-6}}
\tablewidth{0pt}
\tablehead{
\colhead{Julian Date} & \colhead{$I$} & \colhead{$\sigma$} & \colhead{Observatory} }

\startdata

 2450017.8375  &   14.449  &  0.011  &   VVO\\
 2450021.7443  &   14.521  &  0.024  &   VVO\\
 2450026.7416  &   17.132  &  0.074  &   VVO\\
 2450028.6983  &   14.687  &  0.014  &   VVO\\
 2450028.8949  &   14.435  &  0.015  &   VVO\\
 2450031.7696  &   14.014  &  0.007  &   VVO\\
 2450039.7466  &   14.562  &  0.010  &   VVO\\
 2450054.6473  &   14.466  &  0.009  &   VVO\\
 2450056.7206  &   14.466  &  0.007  &   VVO\\
 2450056.8106  &   14.457  &  0.009  &   VVO\\

\enddata

\tablenotetext{a}{The full version of this table is available in electronic
format.  Units of measurements are magnitudes.}

\end{deluxetable}

\clearpage
\begin{deluxetable}{cccccc}
\tablecaption{Color Measurements of KH~15D.\tablenotemark{a}\label{tbl-7}}
\tablewidth{0pt}
\tablehead{
\colhead{$V-I$} & \colhead{$\sigma$} & \colhead{$I$} & \colhead{$\sigma$} & \colhead{Avg. JD} & \colhead{Observatory} }

\startdata

1.619  &   0.037  &  14.475  &   0.024  &  2452237.7643  &  USNO\\
1.582  &   0.024  &  14.525  &   0.018  &  2452241.8421  &  USNO\\
1.416  &   0.064  &  18.061  &   0.037  &  2452261.8464  &  KPNO\\
1.379  &   0.040  &  17.833  &   0.026  &  2452262.8831  &  KPNO\\
1.516  &   0.047  &  17.601  &   0.031  &  2452263.8295  &  KPNO\\
1.565  &   0.039  &  16.739  &   0.023  &  2452264.8739  &  KPNO\\
1.584  &   0.013  &  14.695  &   0.009  &  2452266.7740  &  USNO\\
1.621  &   0.014  &  14.519  &   0.008  &  2452269.8189  &  USNO\\
1.680  &   0.025  &  14.435  &   0.021  &  2452279.6966  &  USNO\\
1.716  &   0.091  &  17.427  &   0.073  &  2452306.8015  &  USNO\\

\enddata

\tablenotetext{a}{The full version of this table is available in electronic
format. While only $V-I$ is shown here, the $B-V$, $B-R$, $V-R$, and $R-I$
colors are also available.  Units of measurements are magnitudes.}

\end{deluxetable}

\clearpage
\begin{deluxetable}{ll}
\tablecaption{Phases Corresponding to Outside of Eclipse. \label{tbl-8}}
\tablewidth{0pt}
\tablehead{
\colhead{Season} & \colhead{Phases} }

\startdata

1995/1996 & 0.20--0.80\\
1996/1997 & 0.20--0.80\\
1997/1998 & 0.20--0.80\\
1998/1999 & 0.25--0.75\\
1999/2000 & 0.25--0.75\\
2000/2001 & 0.25--0.75\\
2001/2002 & 0.30--0.70\\
2002/2003 & 0.30--0.70\\
2003/2004 & 0.35--0.65 \\

\enddata

\end{deluxetable}

\clearpage
\begin{deluxetable}{llll} 
\tablecaption{Detected Periods Outside of Eclipse. \label{tbl-9}}
\tablewidth{0pt}
\tablehead{
\colhead{Season} & \colhead{Observatory} & \colhead{Period (d)} & \colhead{Power}}

\startdata

2001/2002 & CTIO & 9.613, 7.953 & 14.6, 11.9\\
2003/2004 & Tenagra & 9.619 & 8.7 \\
 
\enddata

\end{deluxetable}

\clearpage
\begin{figure}
\plotone{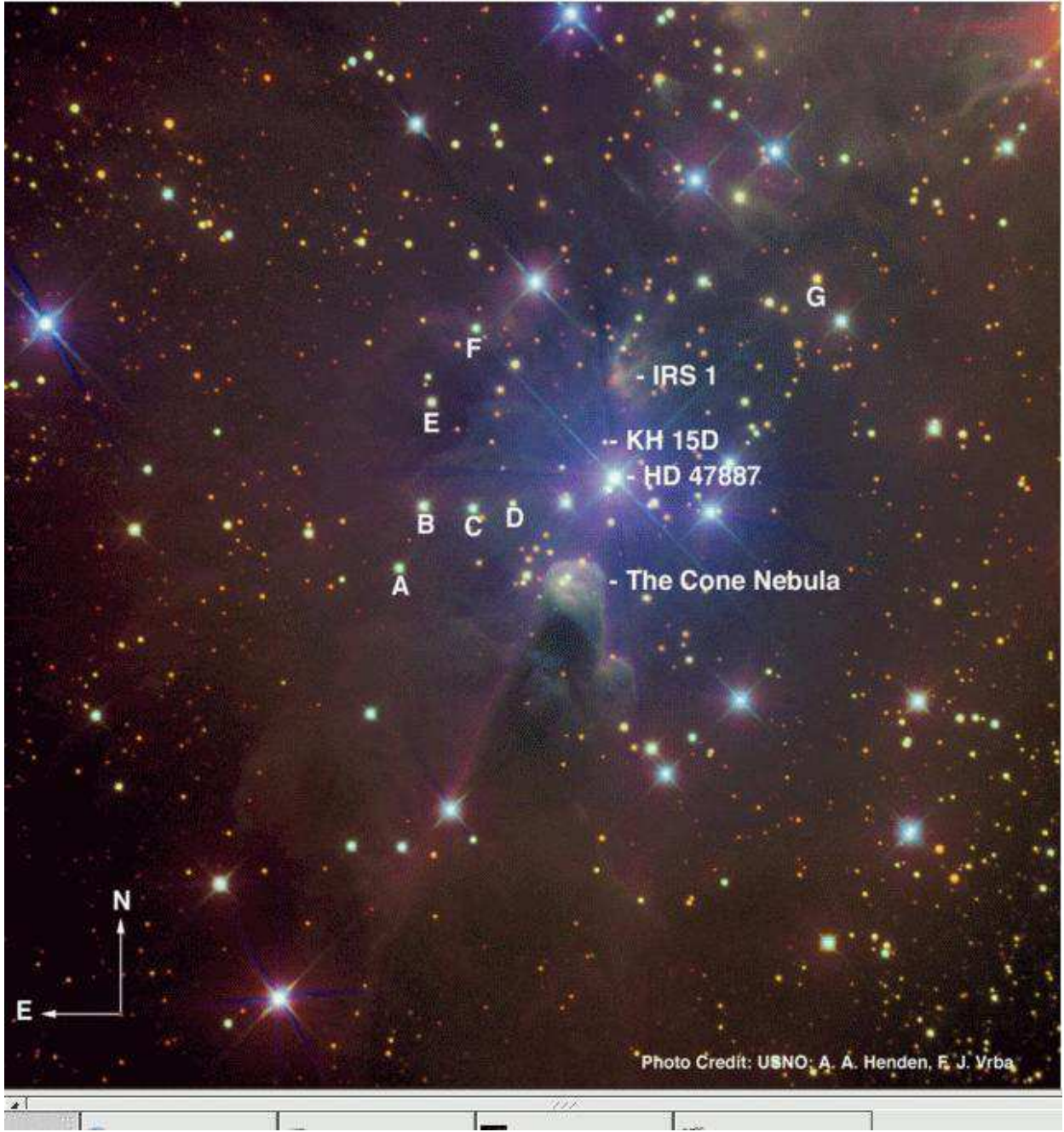}
\figcaption{KH~15D and the surrounding region of NGC 2264.  Local comparison 
stars used in the photometry of KH~15D are labeled A--G.  North is up and
east is to the left.  \label{Fig. 1}}
\end{figure}

\clearpage
\begin{figure}
\plotone{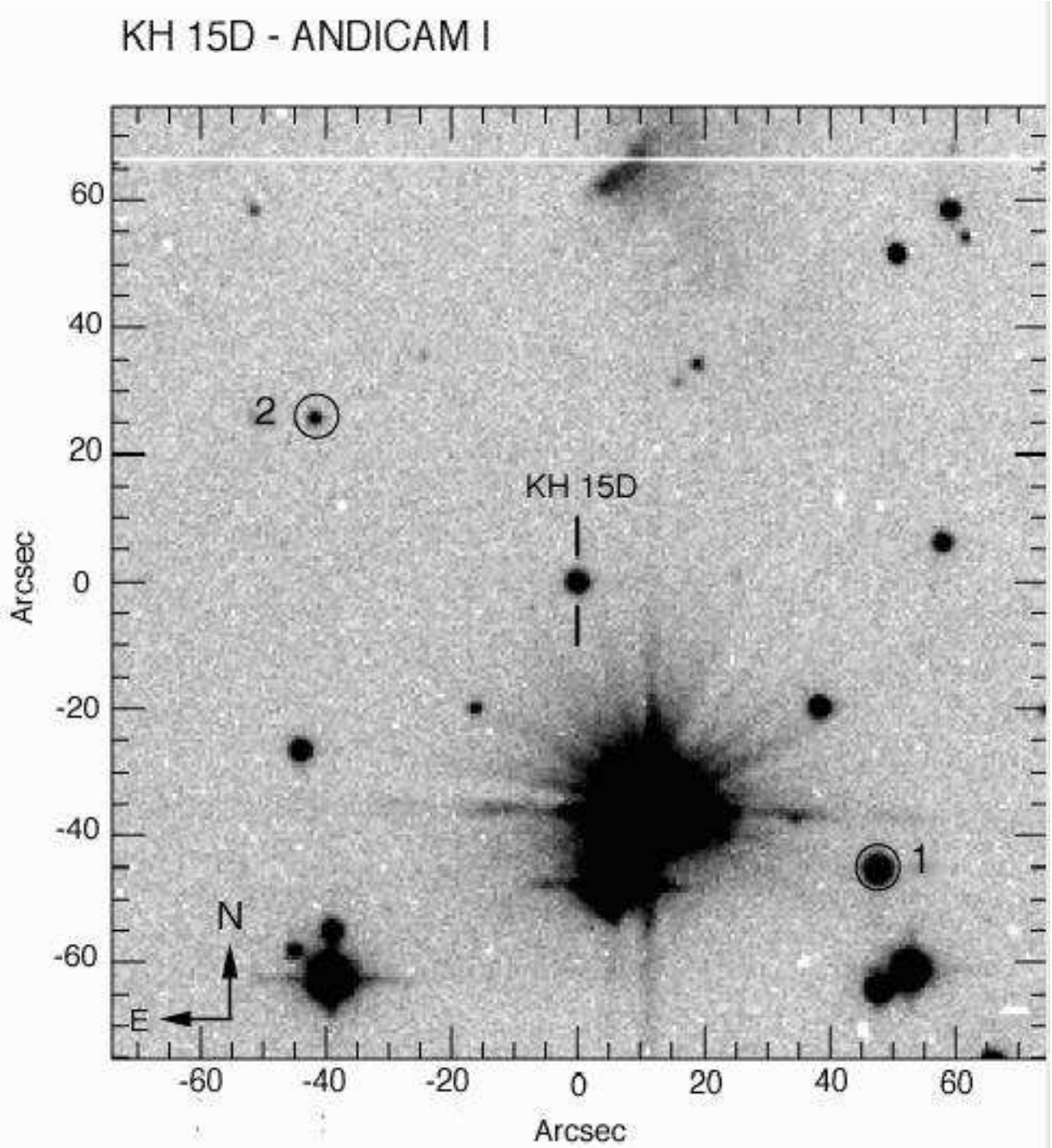}
\figcaption{KH~15D and the local comparison stars used for the 
CTIO photometry. North is up and east is to the left. \label{Fig. 2}}
\end{figure}

\clearpage
\begin{figure}
\plotone{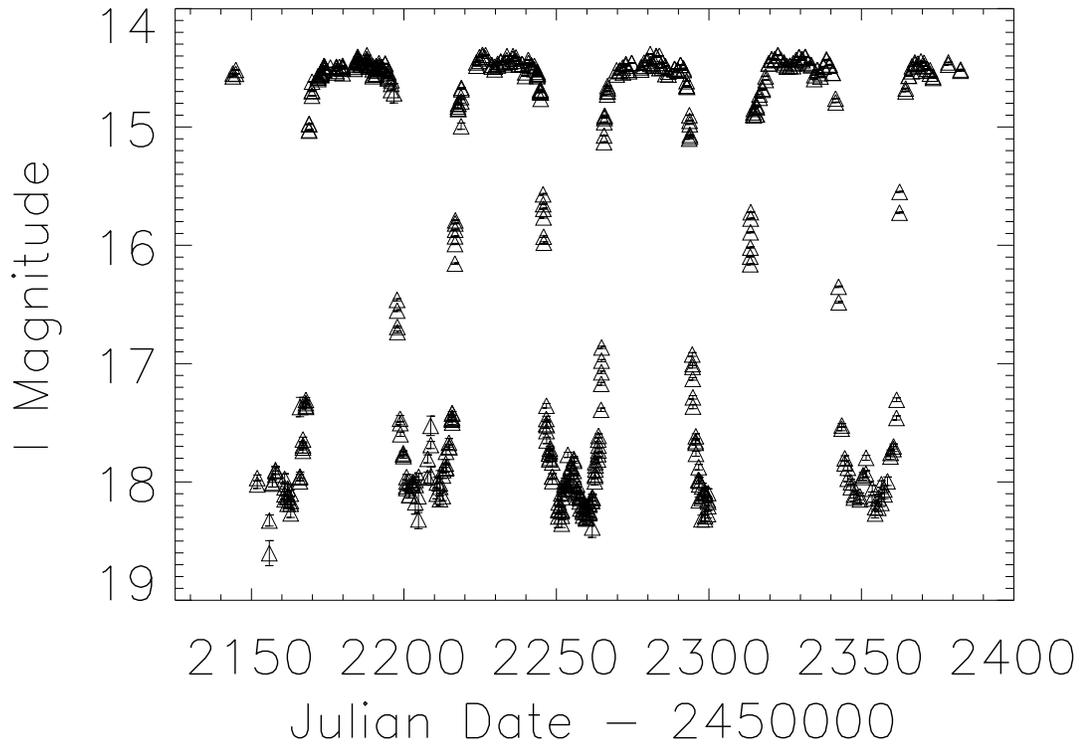}
\figcaption{The light curve of KH~15D during 2001/2002 as observed from CTIO.  
Individual measurements are plotted.  Variability out-of-eclipse
is evident, in addition to changes in the shape of egress from eclipse
to eclipse.  Note the amplitude of the central reversal during this 
season.
\label{Fig. 3}}
\end{figure}

\clearpage
\begin{figure}
\plotone{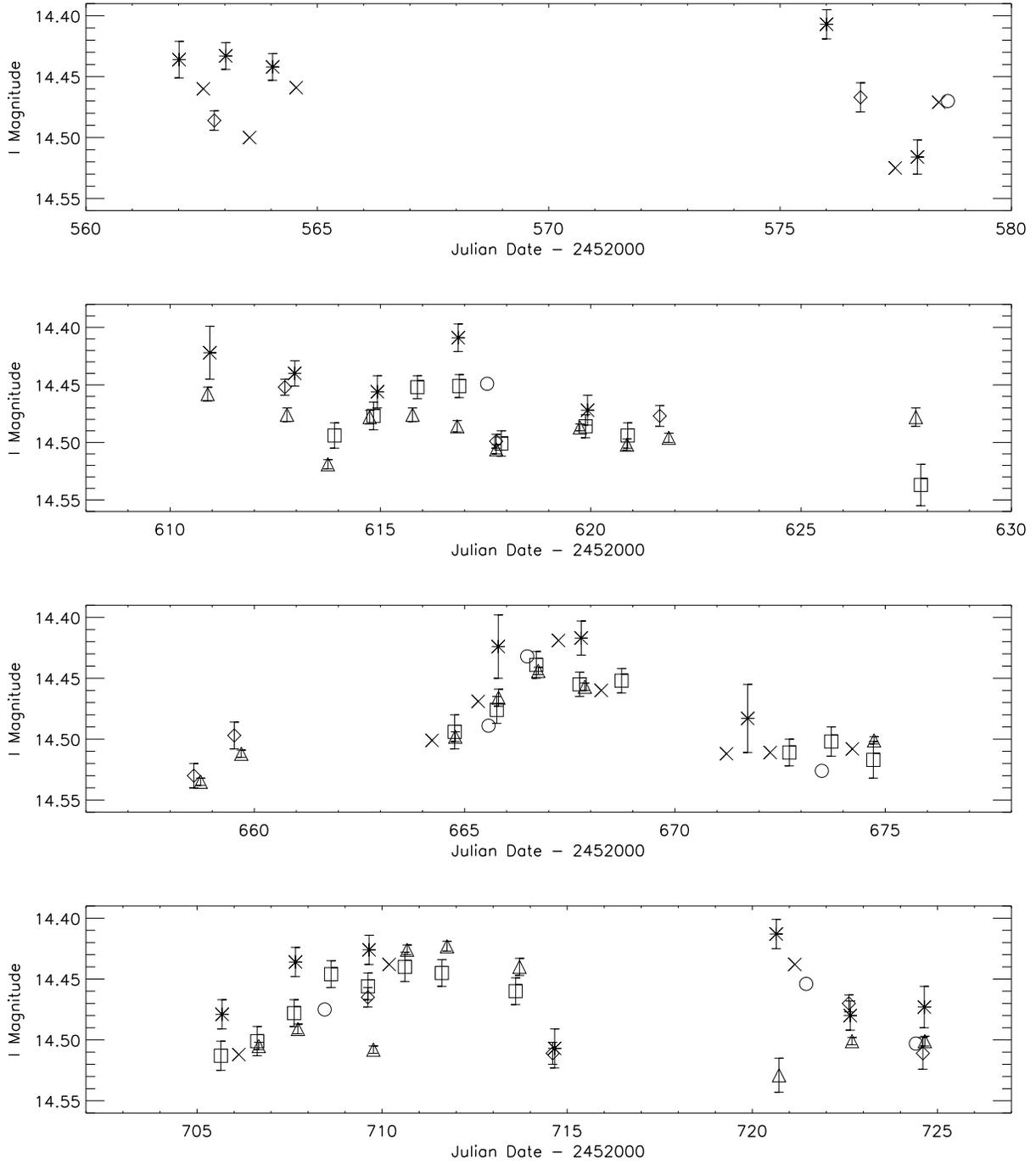}
\figcaption{Data obtained at different observatories on the same 
Julian Dates out-of-eclipse during 2002/2003.
The key to the symbols is as follows: diamonds = VVO, triangles = USNO,
Xs = MMO, squares = Tenagra, asterisks = KAIT,  open
circles = Teide. \label{Fig. 4}}
\end{figure}

\clearpage
\begin{figure}
\plotone{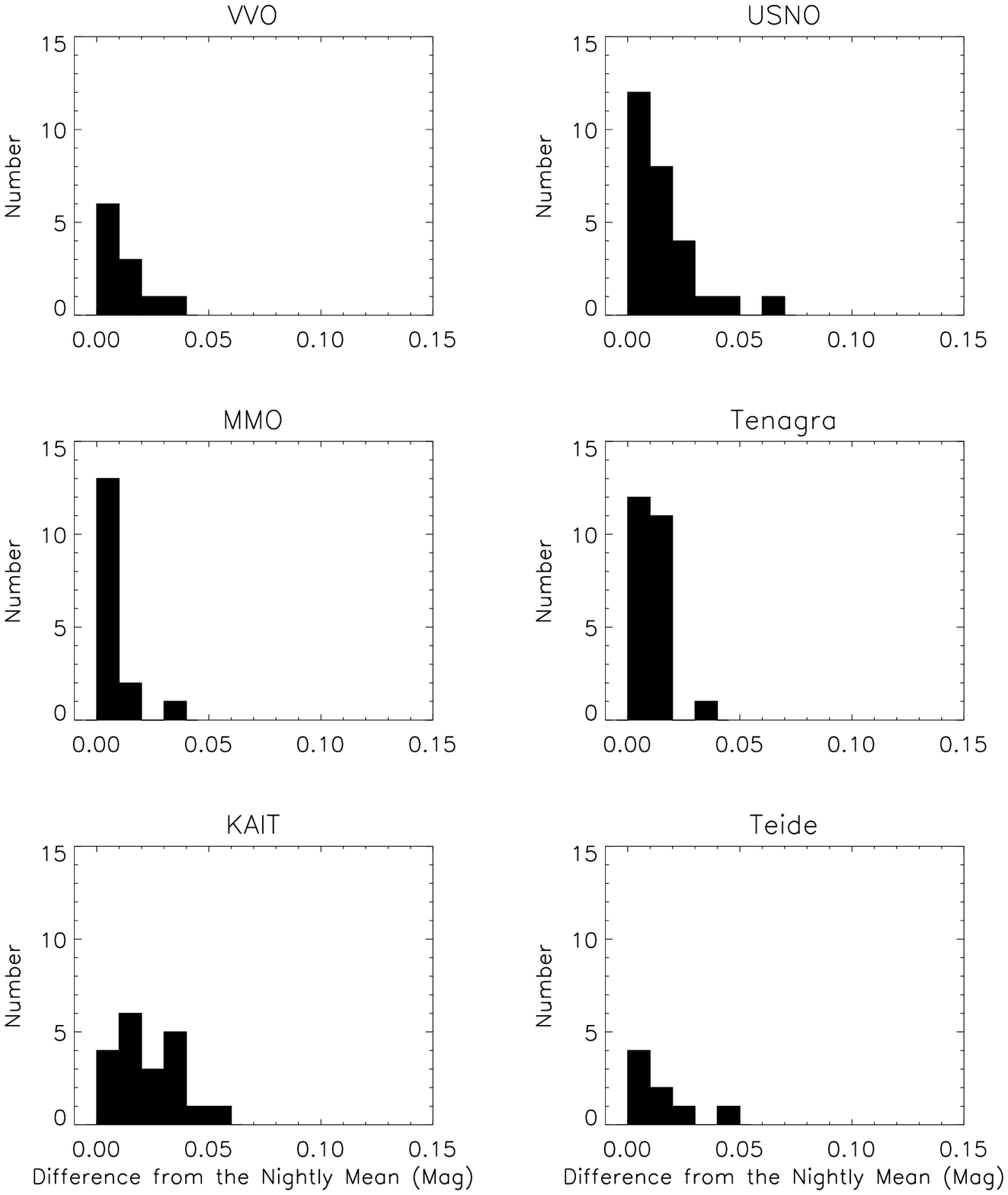}
\figcaption{The number of times an observatory's measurement
differed from the calculated nightly mean out-of-eclipse during 
2002/2003.  
\label{Fig. 5}}
\end{figure}

\clearpage
\begin{figure}
\plotone{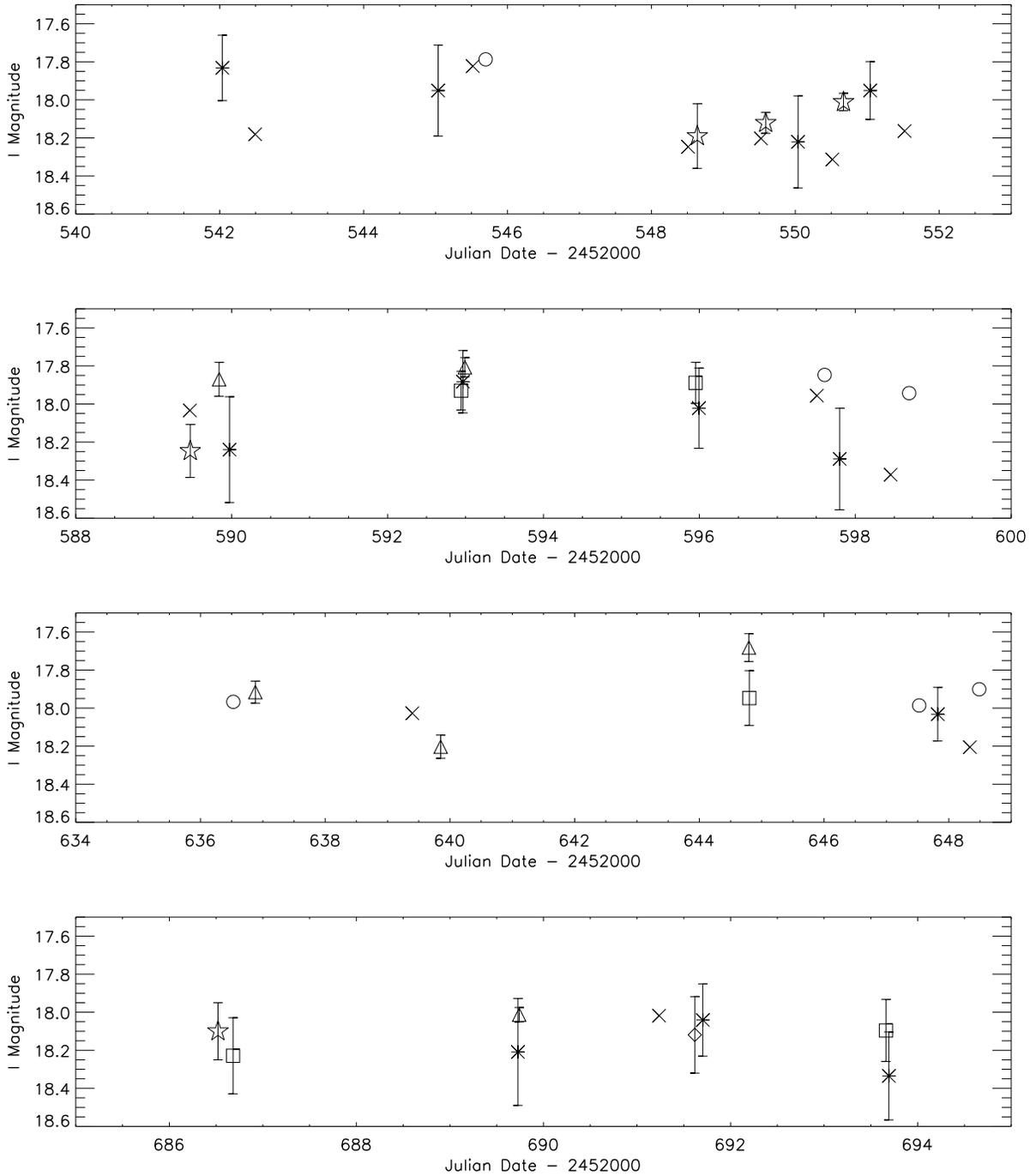}
\figcaption{Data obtained at different observatories on the same 
Julian Dates in deep eclipse during 2002/2003.
The key to the symbols is the same as in Fig. 5: diamonds = VVO, 
triangles = USNO, Xs = MMO, squares = Tenagra, asterisks = KAIT, 
stars = Konkoly, open circles = Teide. \label{Fig. 6}}
\end{figure}

\clearpage
\begin{figure}
\plotone{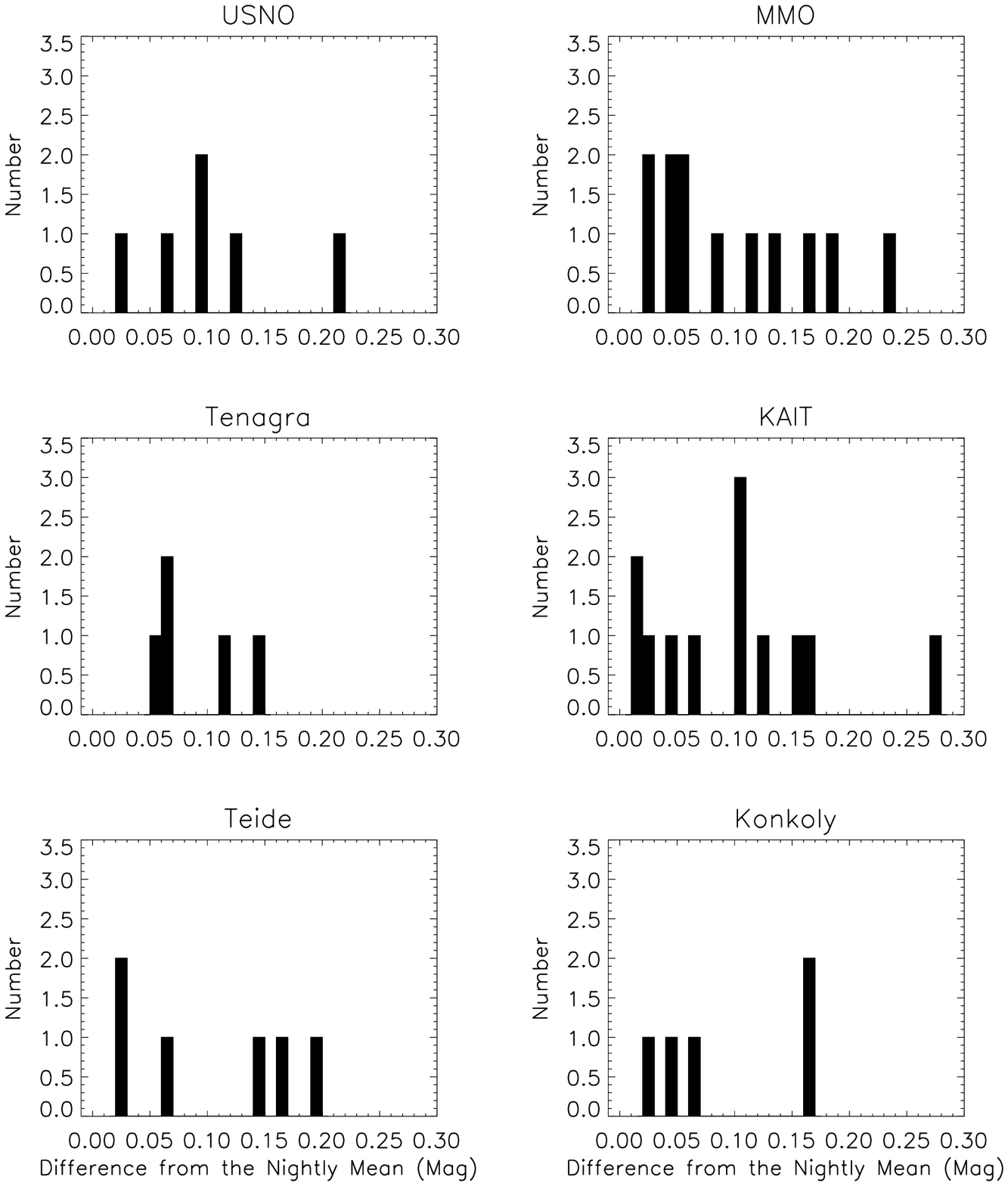}
\figcaption{The number of times an observatory's measurement
differed from the calculated nightly mean in deep eclipse during 2002/2003.
\label{Fig. 7}}
\end{figure}

\clearpage
\begin{figure}
\plotone{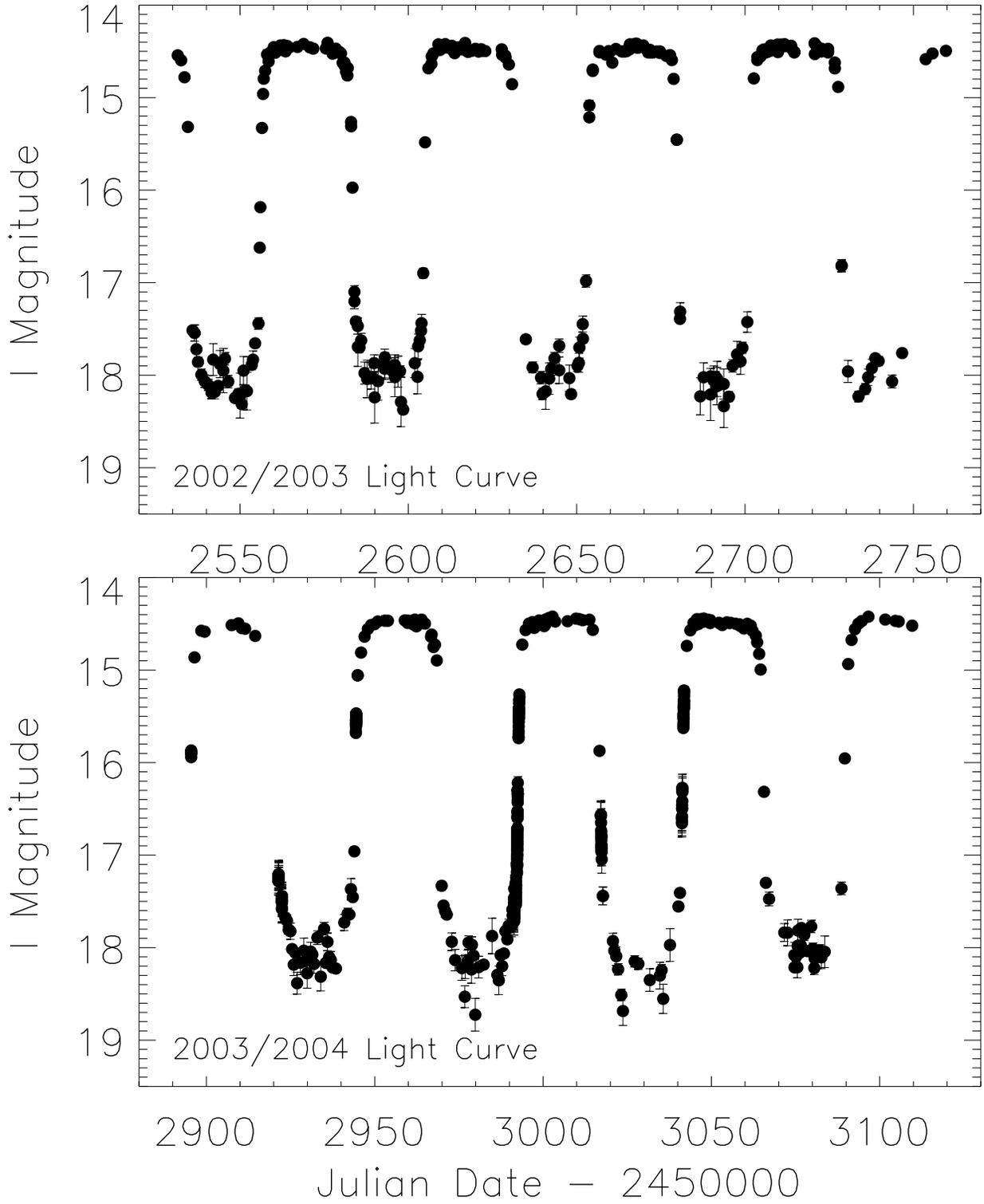}
\figcaption{Light curves from both the 2002/2003 and 2003/2004 seasons.  
The nightly means per observatory are plotted except when rapid changes 
in the brightness are occurring. \label{Fig. 8}}
\end{figure}

\clearpage
\begin{figure}
\plotone{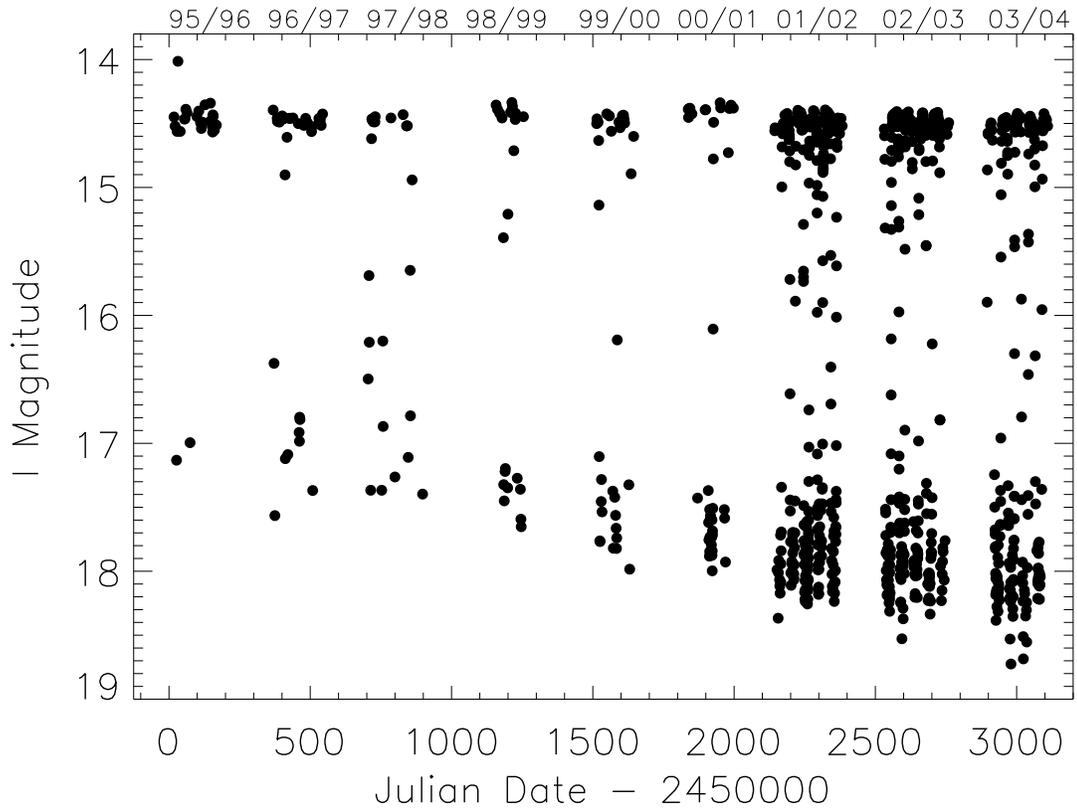}
\figcaption{The light curves of KH~15D over a nine-year
period.  Mean $I$ magnitudes per observatory are plotted.  The data 
span from 1995/1996, the first season of observation
at VVO, to the most recent observing campaign of 2003/2004.  There
appears to be a linear relationship between depth and time.  
\label{Fig. 9}}
\end{figure}

\clearpage
\begin{figure}
\plotone{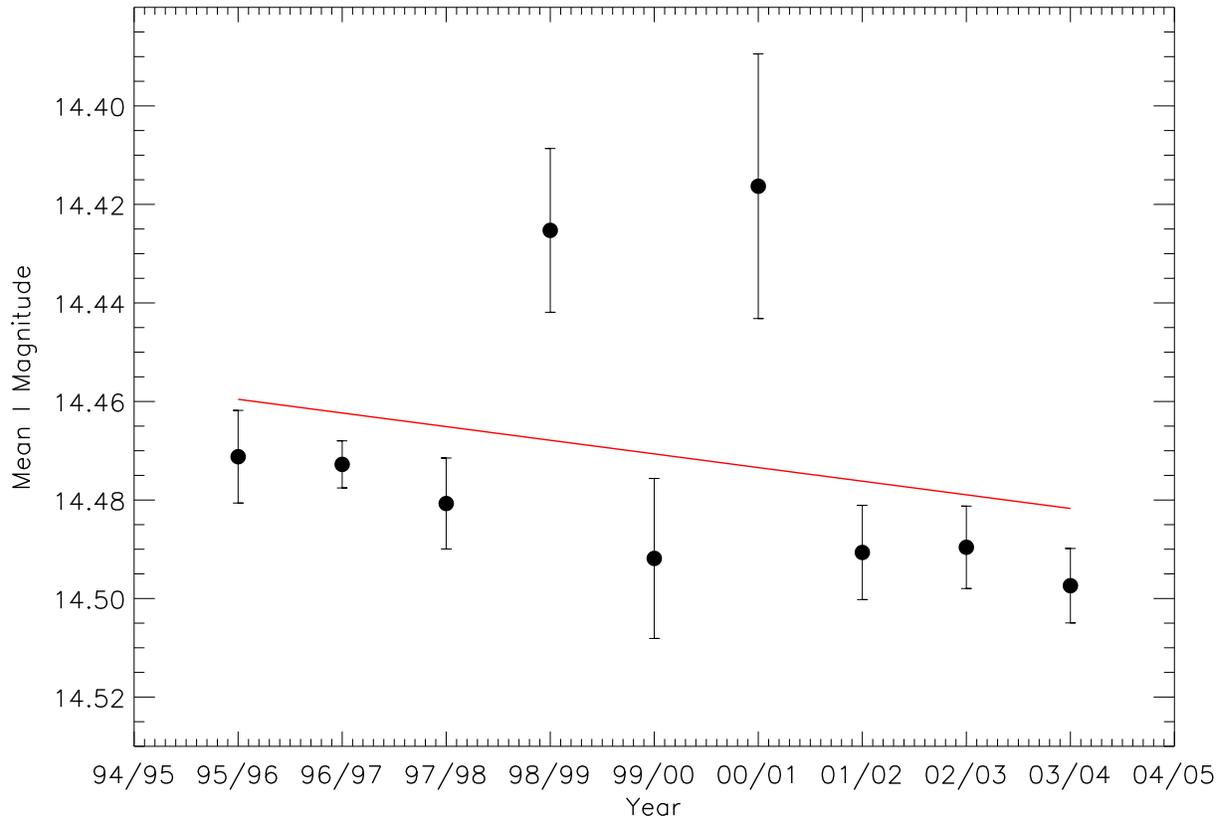}
\figcaption{The mean out-of-eclipse $I$ magnitude as measured at
VVO from 1995 to 2004.  The error bars represent the standard deviation
of the mean for each year.  The straight line is simple linear fit to
the data.  It appears as if the star might be fading a bit with time, but
it is unclear whether this is a significant trend.
\label{Fig. 10}}
\end{figure}

\clearpage
\begin{figure}
\plotone{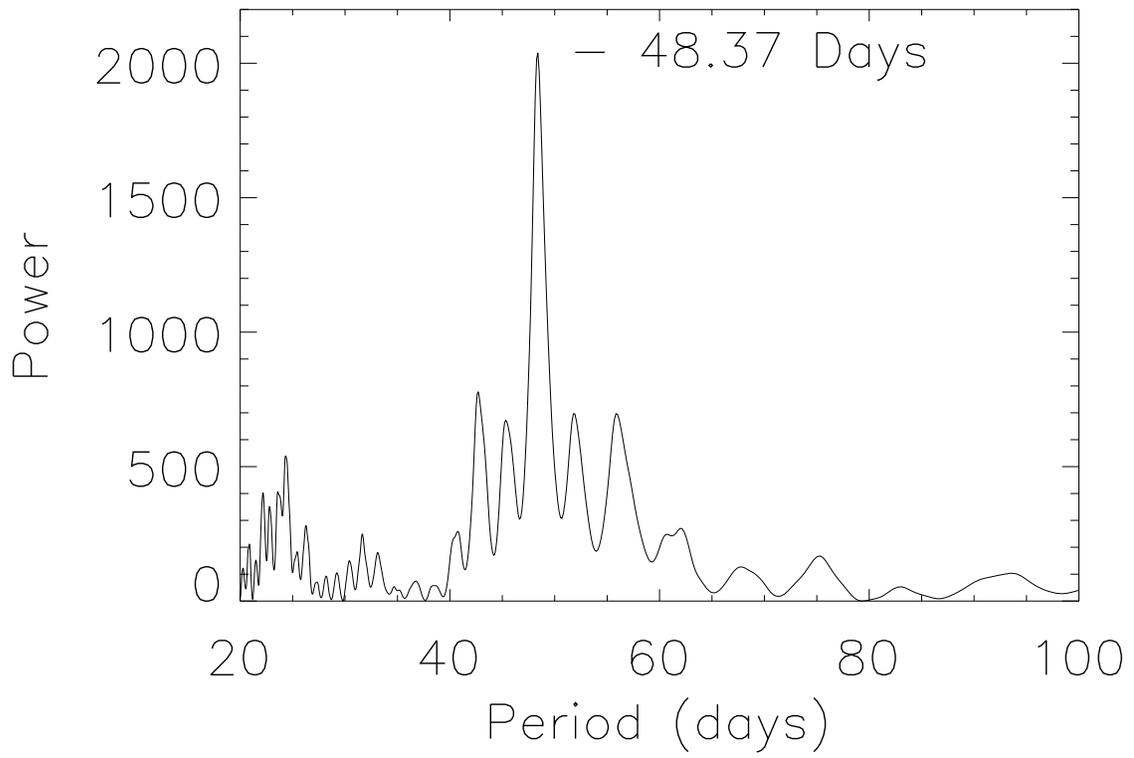}
\figcaption{Power spectrum of all the data during 1995--2004. 
The strongest peak occurs at a period of 48.37 days.  The other four
strong peaks occur at 42.71, 45.32, 51.78, and 55.84 days, respectively. 
\label{Fig. 11}}  
\end{figure}

\clearpage
\begin{figure}
\plotone{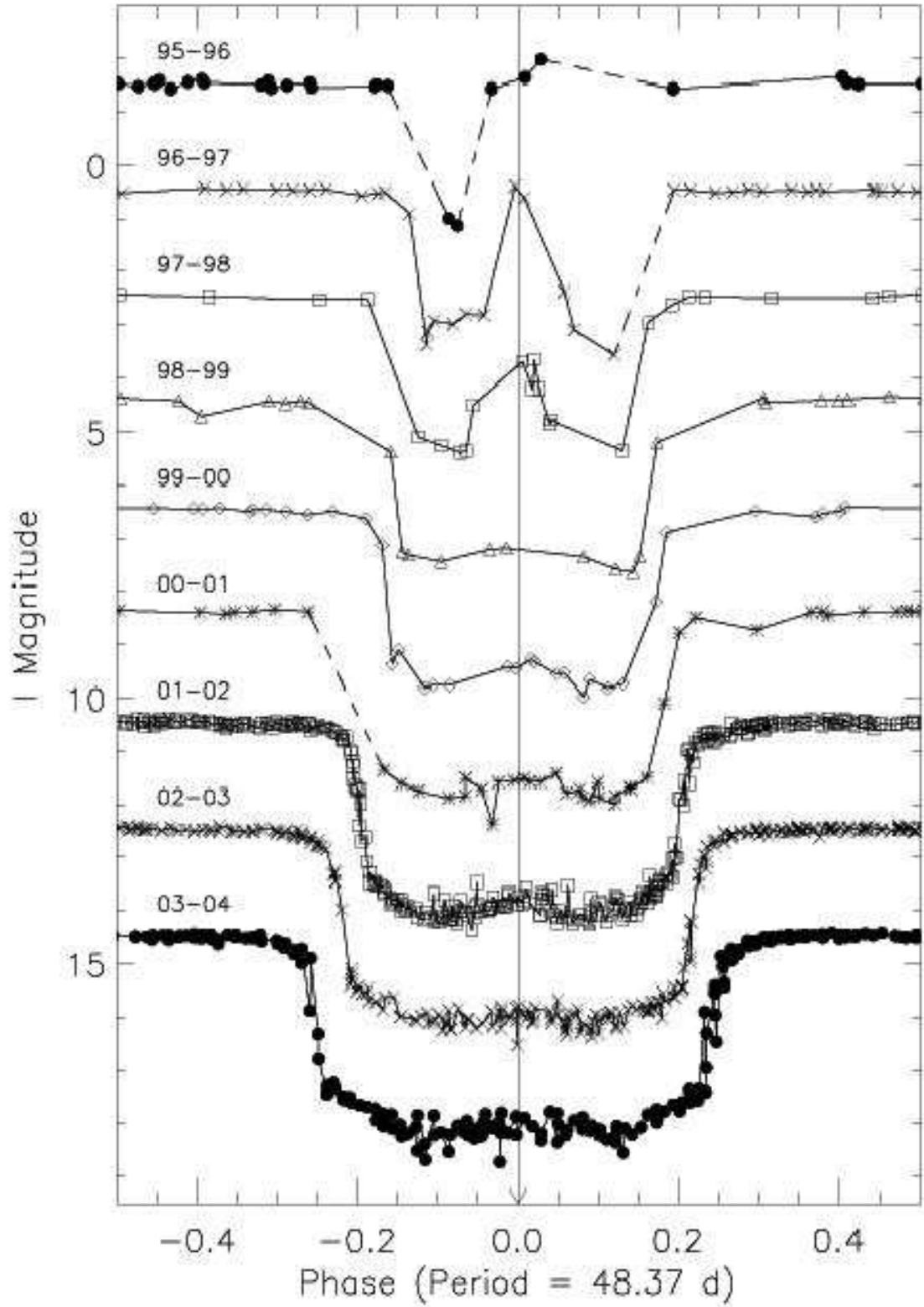}
\figcaption{The data from Fig. 9 are shown here, phased with the 48.37 
day period.  The evolution of the light curve phased at this 
period may tell us something about the dynamical processes that are 
occurring. \label{Fig. 12}}
\end{figure}

\clearpage
\begin{figure}
\plotone{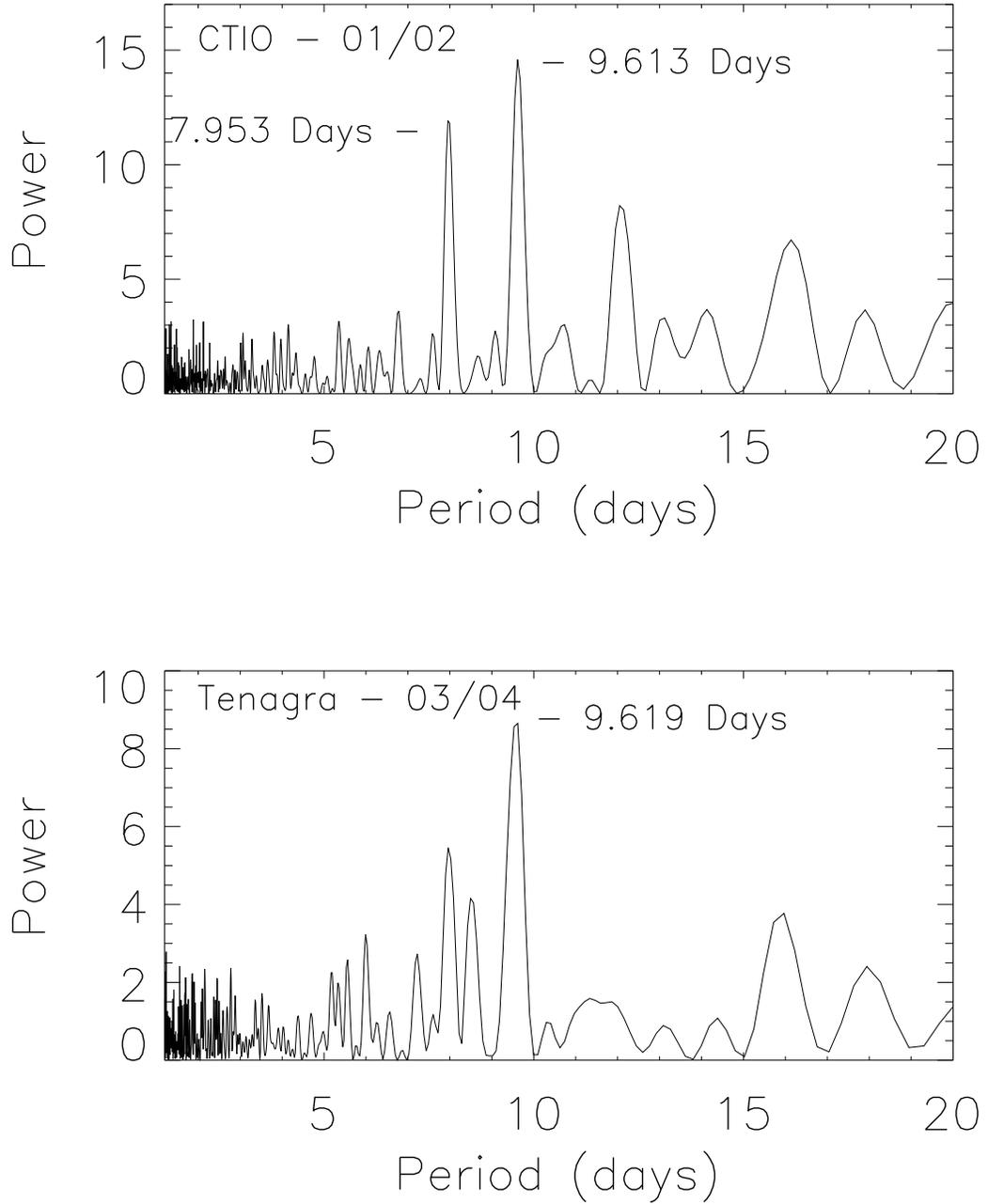}
\figcaption{Power spectra of the out-of-eclipse data obtained at CTIO
during the 2001/2002 season and the out-of-eclipse data obtained at Tenagra 
during the 2003/2004 season.  Each spectrum shows a significant period 
of $\sim$9.6 days. \label{Fig. 13}}
\end{figure}


\clearpage
\begin{figure}
\plotone{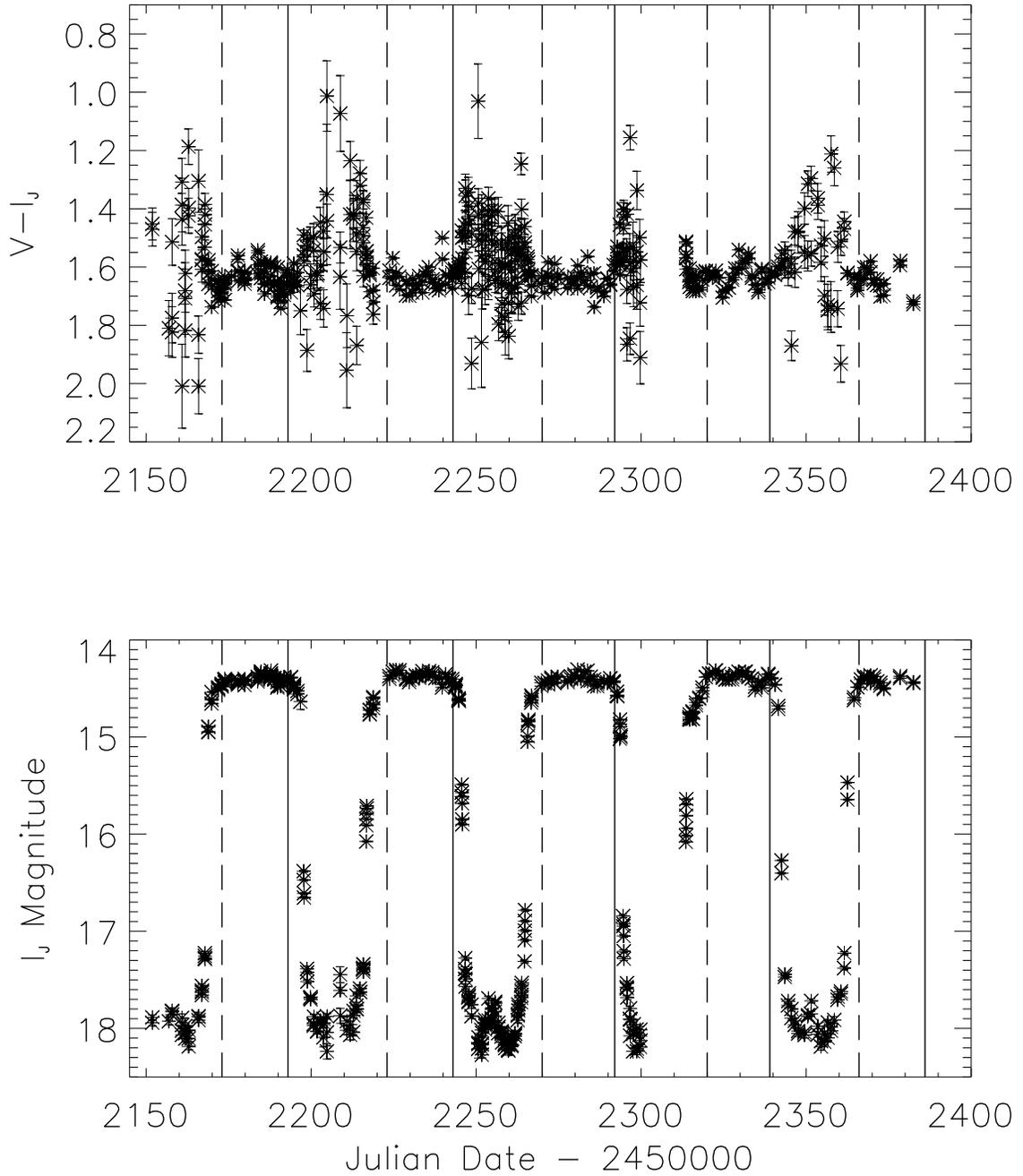}
\figcaption{$V - I_{J}$ color obtained with the CTIO/Yale 1~m telescope 
during 2001/2002.  The solid line represents the approximate beginning
of ingress, whereas the dashed line represents the approximate end of
egress.  The color becomes dramatically bluer during 
ingress, and redder during egress, while it appears to be variable 
throughout the deepest part of the eclipse.
\label{Fig. 14}}
\end{figure}

\clearpage
\begin{figure}
\plotone{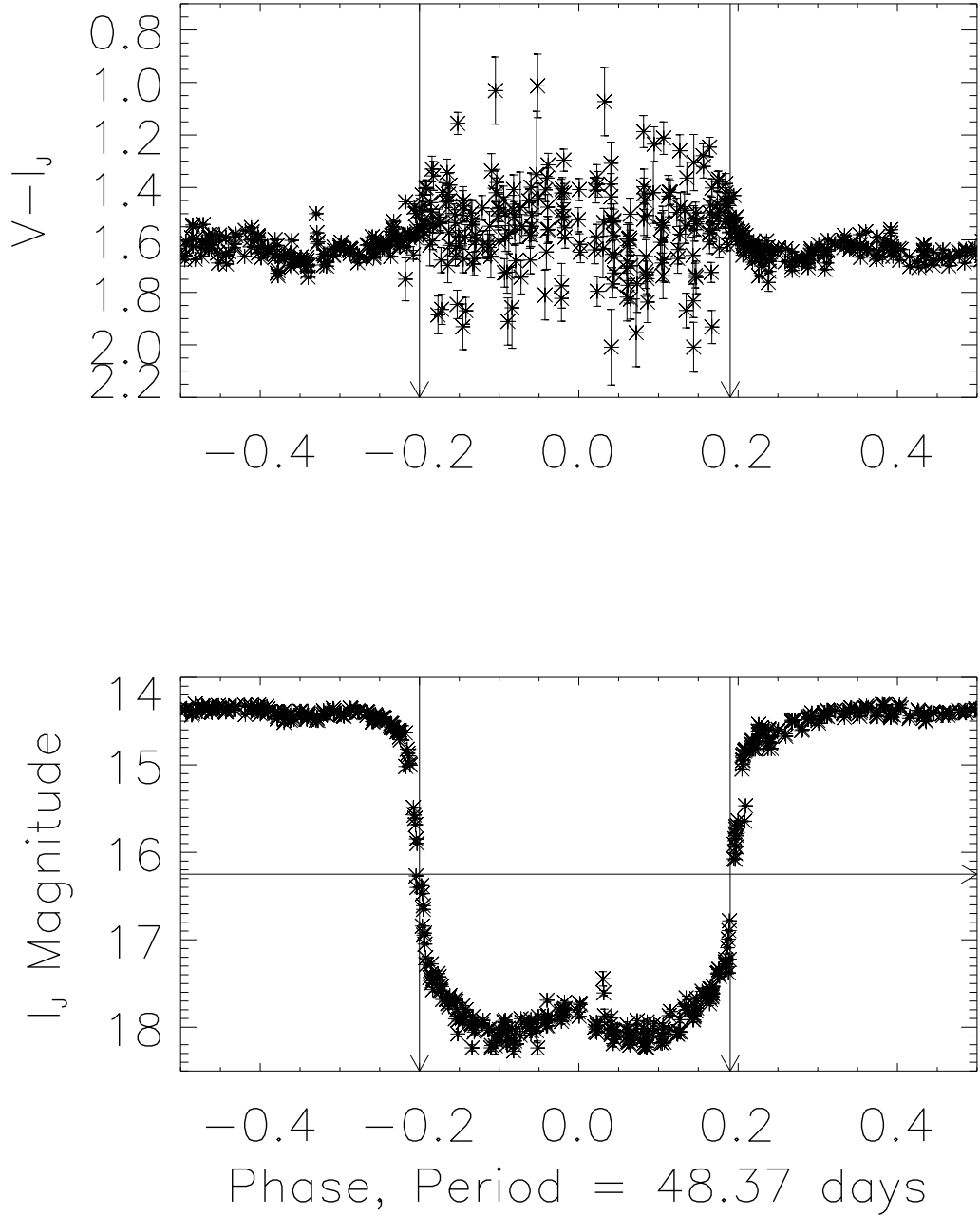}
\figcaption{The data from Fig. 14 are shown here, phased with the 48.37 day 
period.  In this figure, the arrows designate the approximate midpoints
of ingress and egress, as well as where $I$ = 16.25 mag.  The color becomes 
dramatically bluer during ingress, and redder during egress, 
while it appears 
to be variable throughout the deepest part of the eclipse.
\label{Fig. 15}}
\end{figure}

\clearpage
\begin{figure}
\plotone{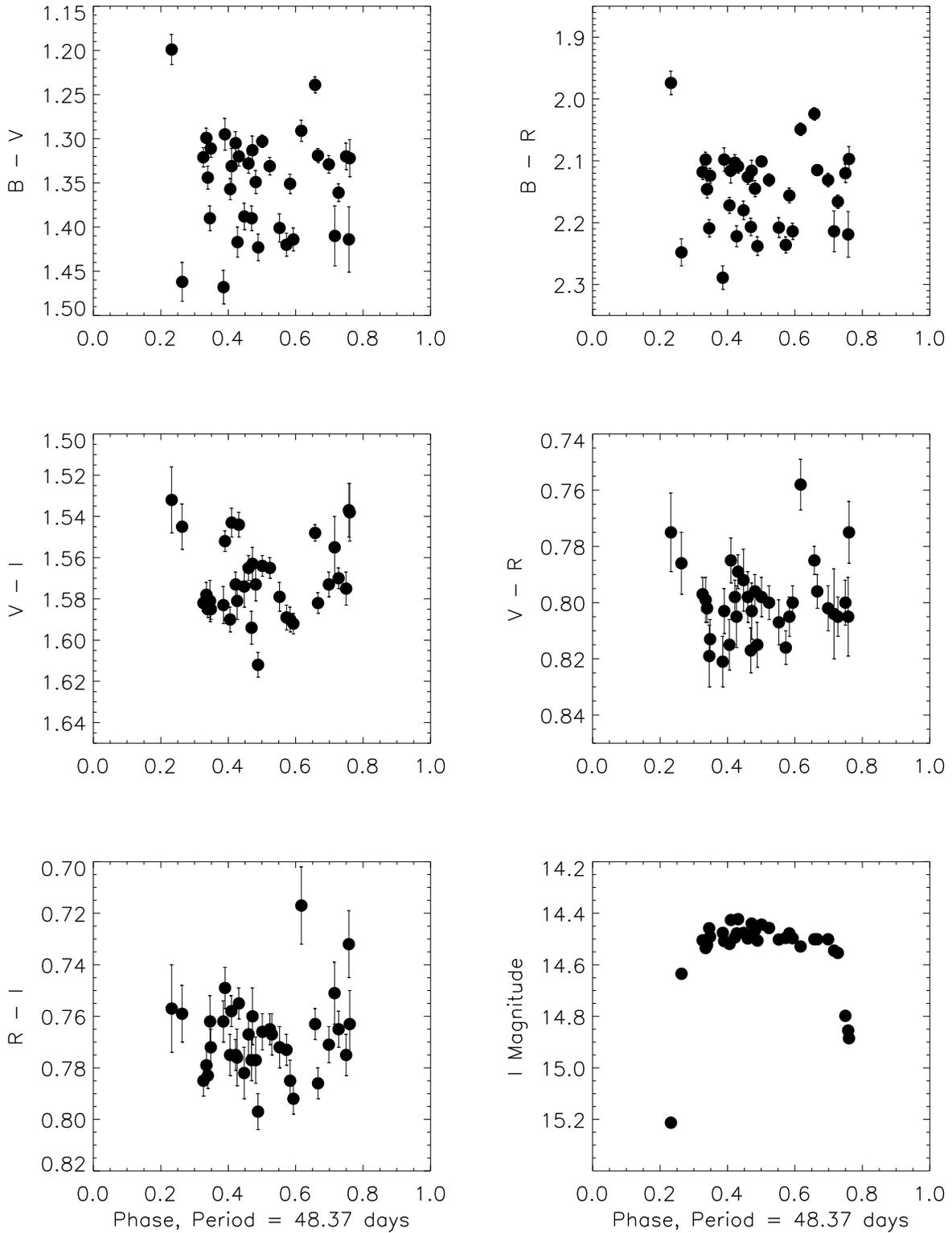}
\figcaption{Color data obtained out-of-eclipse during the 2002/2003
observing season at the USNO, Flagstaff Station versus phase.
The star appears to be irregularly variable in all colors out-of-eclipse.
\label{Fig. 16}}
\end{figure}

\clearpage
\begin{figure}
\plotone{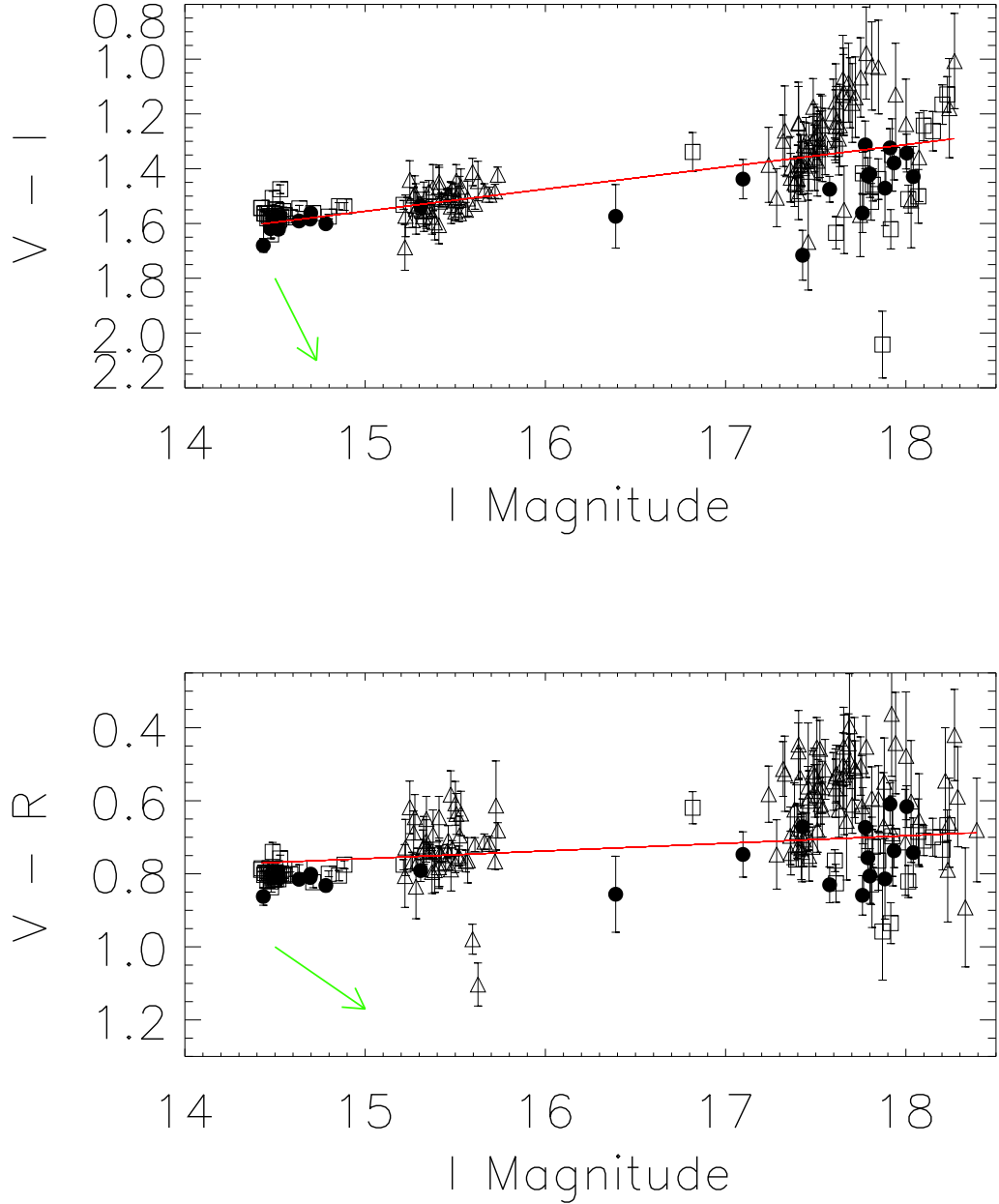}
\figcaption{Here we show all the $V - I$ and $V - R$ data obtained at
the USNO versus $I$ magnitude for
2001--2004 combined. Solid circles represent data from 2001/2002, squares
represent data from 2002/2003, and diamonds represent data from 2003/2004.
A straight line has been fit to the colors, while
the short arrow represents the standard reddening that would be expected
from IS dust grains for each color.  \label{Fig. 17}}
\end{figure}

\clearpage
\begin{figure}
\plotone{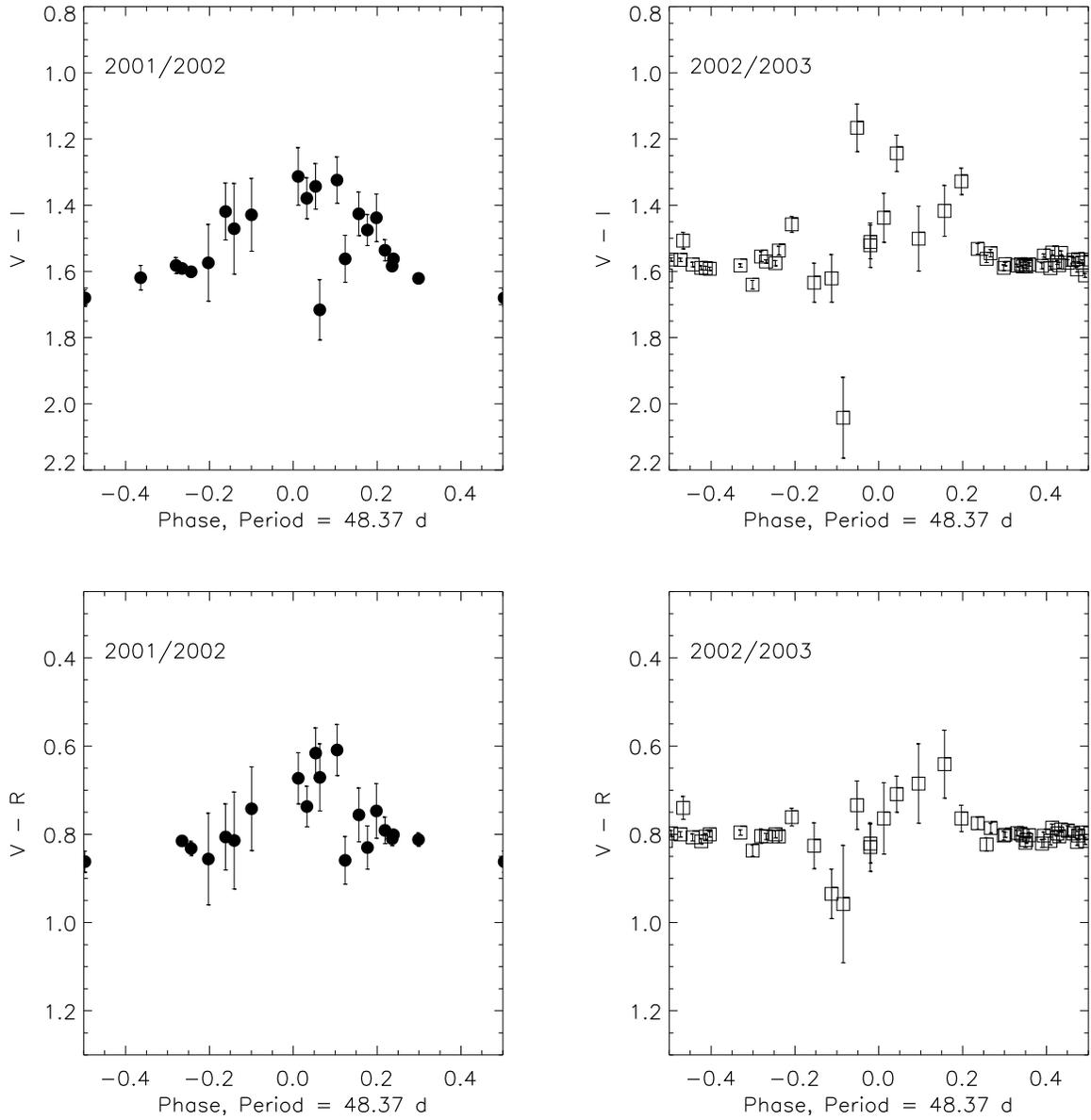}
\figcaption{Here we show all the $V - I$ and $V - R$ data obtained at
the USNO versus phase for
2001--2003. Solid circles represent the $V - I$ data and squares represent
the $V - R$ data.  The trends seen in the CTIO colors are also seen here.
\label{Fig. 18}}
\end{figure}

\end{document}